\documentclass{elsarticle}

\usepackage{subfigure}
\usepackage{graphicx}
\usepackage{verbatim}

\begin{document}
\begin{frontmatter}
\title{Microscopic insights into pedestrian motion through a bottleneck, resolving spatial and temporal variations}

\author[JSC]{Jack Liddle}
\author[JSC,WUP]{Armin Seyfried\corref{cor1}}
\ead{a.seyfried@fz-juelich.de}

\author[JSC]{Bernhard Steffen}
\ead{b.steffen@fz-juelich.de}

\author[WUP]{Wolfram Klingsch}
\ead{klingsch@uni-wuppertal.de}

\author[WUP]{Tobias Rupprecht}
\ead{rupprecht@uni-wuppertal.de}

\author[WUP]{Andreas Winkens}
\ead{winkens@uni-wuppertal.de}

\author[JSC]{Maik Boltes}
\ead{m.boltes@fz-juelich.de}

\cortext[cor1]{Corresponding author}
\address[JSC]{J\"ulich Supercomputing Centre, Forschungszentrum J\"ulich.}
\address[WUP]{Institute for Building Material Technology and Fire Safety, Universit\"at Wuppertal.}
\begin{abstract}
The motion of pedestrians is subject to a wide range of influences and exhibits a rich phenomenology.  To enable precise measurement of the density and velocity we use an alternative definition using Voronoi diagrams which exhibits smaller fluctuations than the standard definitions.  This method permits examination on scales smaller than the pedestrians.  We use this method to investigate the spatial and temporal variation of the observables at bottlenecks.  Experiments were performed with 180 test subjects and a wide range of bottleneck parameters.  The anomalous flow through short bottlenecks and non-stationary states present with narrow bottlenecks are analysed.
\end{abstract}

\end{frontmatter}


\section{Introduction}
The relation between the flow and the bottleneck width and length and the fundamental diagram can be used to validate models and to design facilities.  Many experiments have been performed to study different aspects of pedestrian flow through a bottleneck (for a summary see~\cite{Schadschneider2009a}).  Up to now these studies have considered bottlenecks with different shape, primarily to resolve questions about lane formation, motivation and its effects on the pedestrian flow rate~\cite{Kretz2006,Muir1996,Muller1981,Nagai2006,Seyfried2009,Cepolina2009,Daamen2010,Hoogendoorn2005,Yanagisawa2009,Helbing2005}.  Statements relating the flow to bottleneck parameters should be generally valid and not depend on the initial conditions.  If a system retains information about its initial conditions then it may exhibit a non-stationary state, where quantities such as the density will show a time dependence.  Analysis of stationarity in these experiments is limited by low participation rates.  Effects on stationarity due to the bottleneck width cannot be addressed due to the narrow range of bottleneck widths studied.  Comparison of these experiments is also complicated by the differing initial arrangements of the pedestrians.  Short bottlenecks (to reproduce motion through a doorway) have been studied~\cite{Muller1981,Daamen2010} and an improved flow rate is seen in agreement with~\cite{Liddle2009}, however the former only studied a single bottleneck length so it cannot held to be a conclusive result.

In~\cite{Liddle2009} we observed that the flow exhibited a time dependence (with the flow decreasing with time).  We were able to make such statements because our experiments had a much higher participation level than other experiments which have investigated the flow through a bottleneck.  Some works have partly addressed the stationarity question, through a fitting of an exponential decay to the pedestrian velocities~\cite{Seyfried2009} or time gaps between successive pedestrians leaving the bottleneck~\cite{Kretz2006}.  In the case of velocity fits, participation was not great enough in order to remove transient behaviour.  Due to the massive fluctuations in the time gaps no model can be reliably fitted.

The participants in previous experiments  were mixed across social groups.   Unsurprisingly the mix is not consistent between these experiments, and this poses additional problems when trying to compare the experiments.  In~\cite{Daamen2010} the effect of population composition on the flow rate through door ways is considered.

If non-stationarity is to be detected a quantity whose fluctuations permit the detection of a trend in the data over the course of the experiment is needed.  Standard definitions of the density exhibit large fluctuations as typically the size of the measurement area is comparable with the size of pedestrians and the size of the measurement area must be larger than the pedestrians.  This limits the spatial resolution and stationary states cannot be unambiguously detected.  An alternative definition using Voronoi diagrams exhibits smaller fluctuations exposing non-stationary in our data.  This method also permits examination on scales smaller than the pedestrians.

\begin{figure}[h]
      \centering
      \subfigure[Combined stills taken from experiment.]
      {
         \includegraphics[scale=0.35]{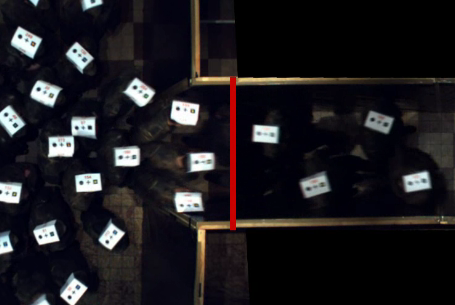}
         \label{fig:experimental-still}
      }
      \subfigure[Experimental setup.]
      {
         \includegraphics[scale=0.42]{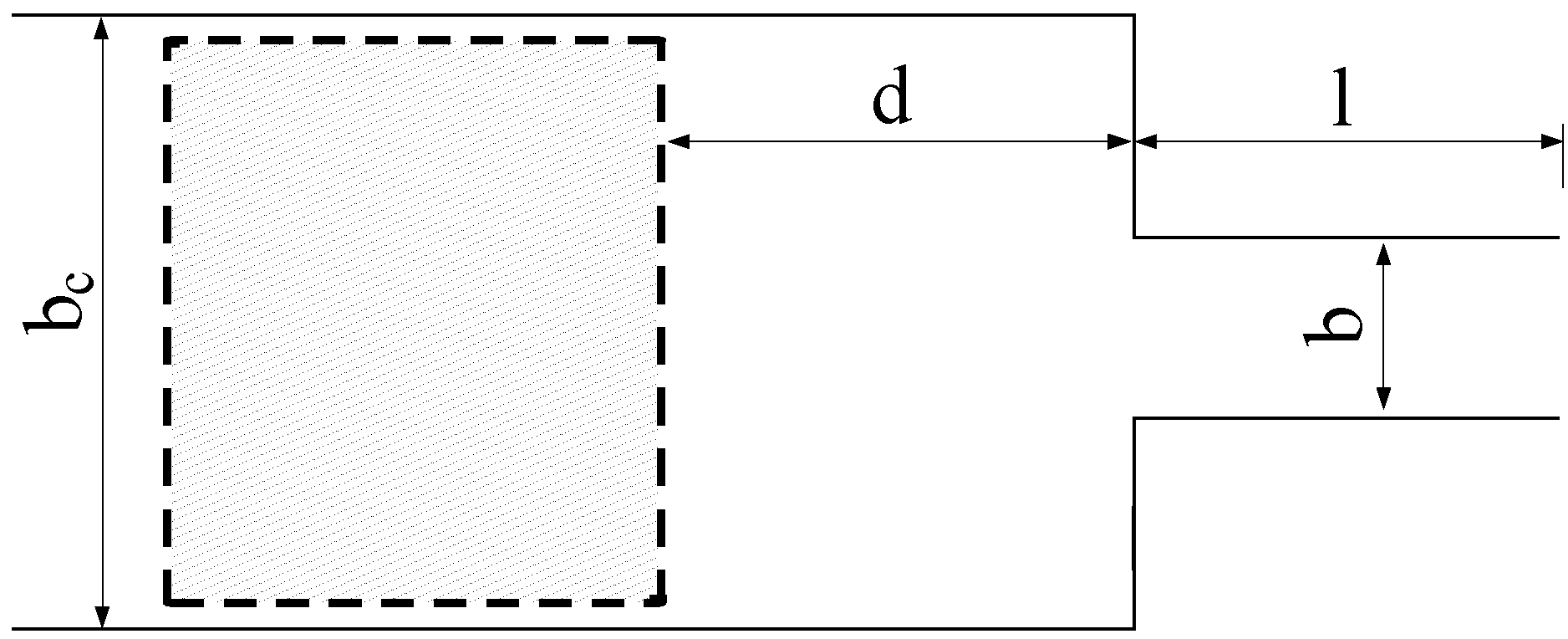}
         \label{fig:experimental-setup}
      }
      \caption{The experimental setup, used in the Bergische Kaserne.}
   \end{figure}

Our experiment was performed in 2006 in the wardroom of the ‘Bergische Kaserne D\"usseldorf’, with a test group of soldiers.  The experimental setup allowed the influence of the bottleneck width and length to be probed.  In one experiment the width, $b$, was varied (from $90$ to $250$ cm) at a fixed corridor length of 400 cm, another experiment investigated varying the corridor length, $l$, ($6,200,400$ cm) at a fixed width of $120$ cm.  Wider bottlenecks with more test persons were studied than in previous attempts.  The pedestrians begun their motion from a waiting area, at a distance $d$ from the bottleneck, to ensure an equal initial density for every run, holding areas were marked on the floor (dashed regions).  The density in this waiting area was 2.6 $P/m^2$.  Figure~\ref{fig:experimental-setup} shows a sketch of the experimental setup used to analyze the flow through bottlenecks, figure \ref{fig:experimental-still} shows a still taken from the experiment.  The experiments investigating the bottleneck width took place on a single day, in order of decreasing bottleneck width.  After every 5 experiments a pause was taken.  No systematic qualitative trends were noted during the course of the experiments.  The following day the length variation experiments took place in a single run, from longest to shortest.

Our series of experiments have several advantages.  The social homogeneity of the test subjects presents a consistent picture of flow characteristics, and the high number of participants will expose any trends in the data.  Only one initial arrangement of the pedestrians is used.  The study of 3 different lengths (6 cm, 200 cm and 400 cm) allows us to address the problem of stationarity and flow through a short bottleneck consistently.  The participation level in our experiment was higher than any previous experiments, combined with the reduced fluctuations inherent to integrated method our experiment has the possibility to detect non-stationarities in the density.  The wider range of bottleneck widths studied allow us to address any effects of bottleneck width on stationarity.

\section{Definitions}

The density has an underlying distribution, $f(\vec{x},t)$, which depends on position and time.  The exact form of $f(\vec{x},t)$ will depend on the measurement method and will exhibit fluctuations in time as the system evolves and in space on the scale of the individual pedestrians.  From the underlying distribution we can obtain an average measurement,
\begin{equation}
   \bar{\rho} = \frac{1}{A\Delta T}\int_{A} \int_{\Delta T} f(\vec{x},t) dt d\vec{x} ,
\end{equation}
over a measurement area, $A$, and over a time interval, $\Delta T$.  Varying the size of the measurement area or length of measurement time interval, will affect the reliability of the density estimate $\bar{\rho}$.  For example if the measurement area is reduced we must use a longer time period in order to obtain a reliable estimate for $\bar{\rho}$, the converse also holds.

The standard definition of the density in a measurement area $A$, of area $|A|$, containing $N$ pedestrians,
\begin{equation}
   \rho_S = \frac{N}{|A|},
\end{equation}
exhibits large fluctuations as pedestrians enter and leave the measurement area, when the area is of the same order as the size of the pedestrians.  These fluctuations are typically of the same order as the density measurement itself.  By averaging over time these fluctuations can be removed at the expense of reduced time resolution.  The underlying density distribution, $f(x,t)$, is a sum of delta functions so it can only be applied on scales larger than the pedestrians, limiting the potential spatial resolution.

\begin{figure}
      \centering
      \subfigure[$b = 120$ cm]
      {
         \includegraphics[scale=0.28]{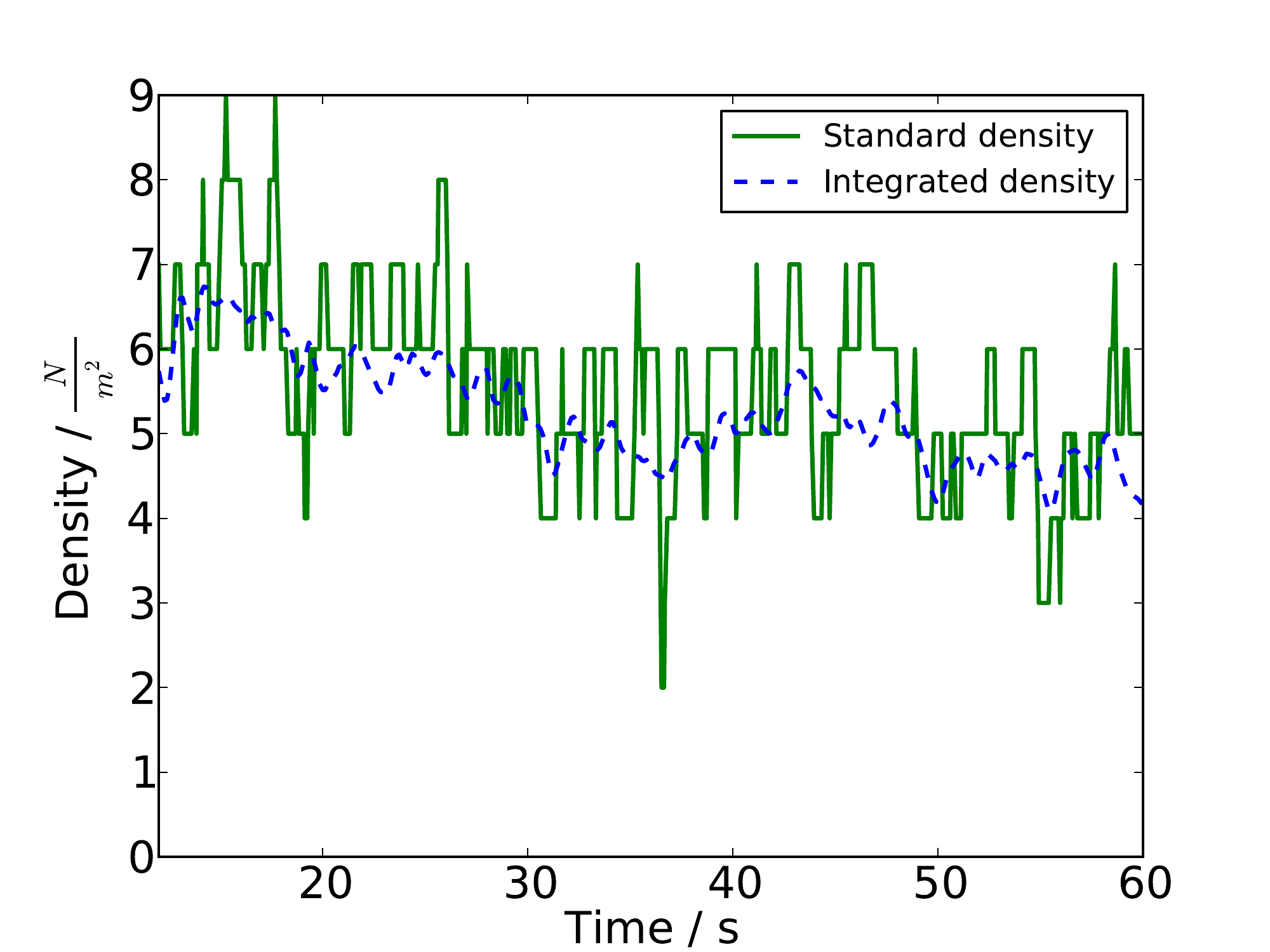}
      }
      \subfigure[$b = 140$ cm]
      {
         \includegraphics[scale=0.28]{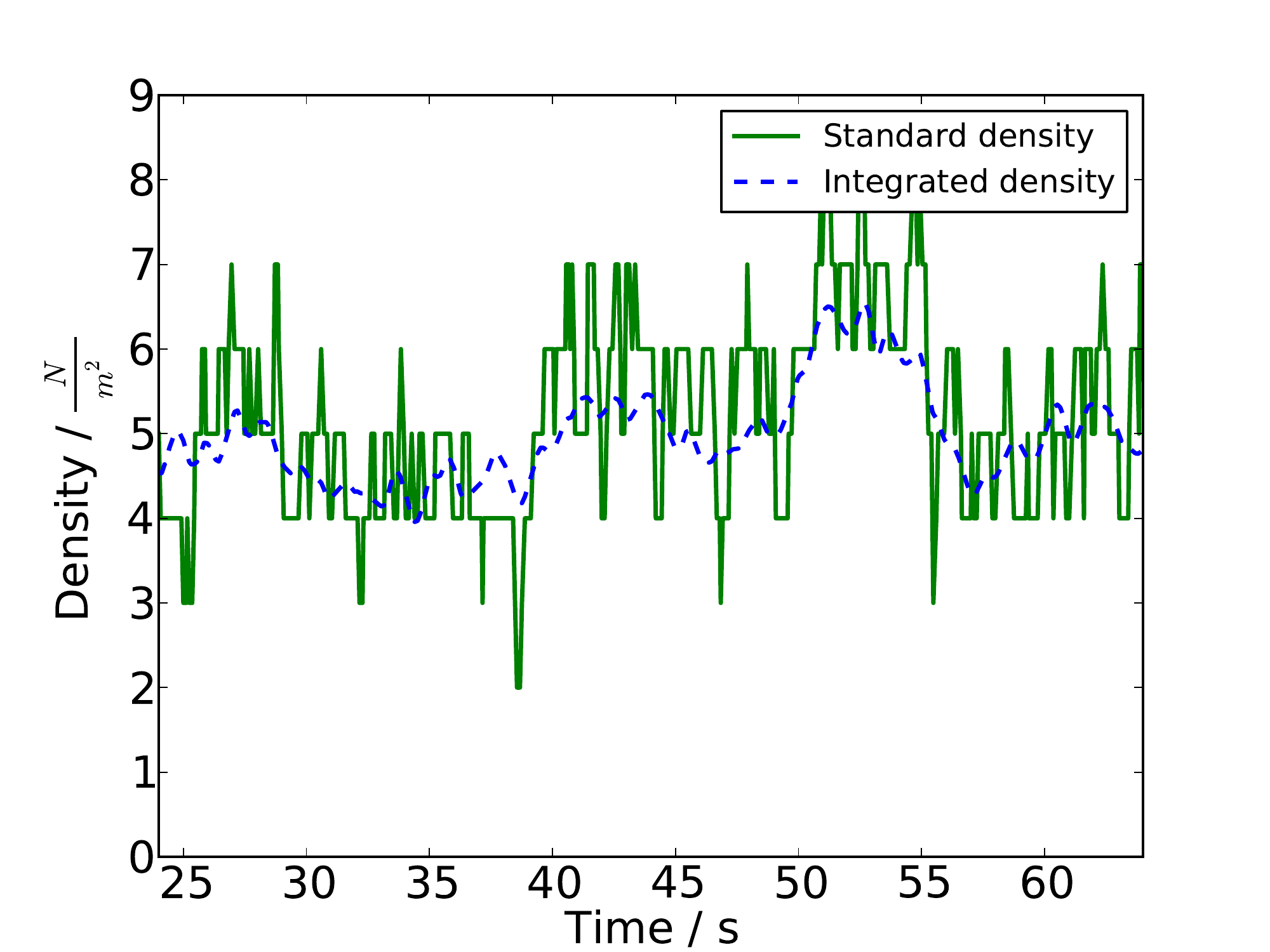}
      }

      \caption{Density computed using standard and integrated definitions.  It can be clearly seen that the integrated definition fluctuates less than the standard definition.  The non-stationarity of the integrated density can be clearly seen ($b = 120$ cm).  The bottleneck is 400 cm long.  The measurement area is 100 cm $\times$ 100 cm, and is centered 100 cm from the bottleneck entrance.}
      \label{fig:density}
   \end{figure}

The focus of this paper is the integrated density method proposed in~\cite{Steffen2010a} where a density distribution (step function) is assigned to each pedestrian.  This density distribution is calculated through the Voronoi diagram.  At a given time, $t_i$, we have a set of positions for each pedestrian $\{x_1(t_i),x_2(t_i),\ldots,x_M(t_i)\}$.  We compute the Voronoi diagram for these points obtaining a set of cells, $A_i$, for each pedestrian $i$.  These cells can be thought of as the \emph{personal} space belonging to each pedestrian.  With these cells a density distribution can be defined for each pedestrian,
\begin{equation}
f_i(x,t) = \left\{
            \begin{array}{ll}
                       \frac{1}{|A_i|} & \mbox{if } x \in A_i \\
                        0 & \mbox{otherwise.}
            \end{array}
            \right.
\end{equation}
The density distribution is the sum of the individual distributions,
\begin{equation}
   p(x,t) = \sum_i f_i(x,t).
\end{equation}
The density over an area $A$ is defined as,
\begin{equation}
   \rho_V(t) = \frac{\int_A p(x,t) dx}{|A|}.
\end{equation}
The integrated method has greatly reduced spatial fluctuations and does not require a large $\Delta T$, to obtain a reliable estimate.  The integrated density also permits examination on scales much smaller than individual pedestrians, at the expense of reduced temporal resolution.  In figure~\ref{fig:density} the density is shown, calculated using both methods.  The integrated density shows much smaller fluctuations than the classical definition.  This method provides several advantages. The reduced fluctuations mean an instantaneous estimate of the density is possible and the presence of non-stationary states can be unambiguously detected, something which is not possible with the standard method, see figure 1. The integrated density can also provide microscopic information about the density, which will inform the development of microscopic models.

The pedestrian velocity is defined as,
\begin{equation}
v_i(t) = \frac{x_i(t+\Delta t) - x_i(t - \Delta t)}{2 \Delta t}.
\end{equation}
The time difference has been chosen to be $\Delta t = 0.2$ (corresponding to 5 frames in advance/behind, for our videos recorded at 25 frames per second).  The \emph{standard velocity} is given by
\begin{equation}
v_S(t) = \frac{\sum_i v_i(t)}{N}.
\end{equation}
Analogous to the integrated density an integrated velocity can be defined.  Individual contributions to the velocity are
\begin{equation}
\nu_i(x) = \left\{
            \begin{array}{ll}
                       \frac{v_i}{|A|} & \mbox{if } x \in A_i \\
                        0 & \mbox{otherwise}
            \end{array}
            \right.
\end{equation}
using the velocity distribution
\begin{equation}
   v(x) = \sum_i \nu_i(x)
\end{equation}
we can obtain the 
\begin{equation}
   v_V = \frac{\int_A v(x) dx}{|A|}
\end{equation}
The magnitude of the hydrodynamic flow,
\begin{equation}
   |J_s| = \rho |v|,
\end{equation}
can also be calculated using the integrated density and velocity.

\begin{figure}
   \centering
   \includegraphics[scale=0.50]{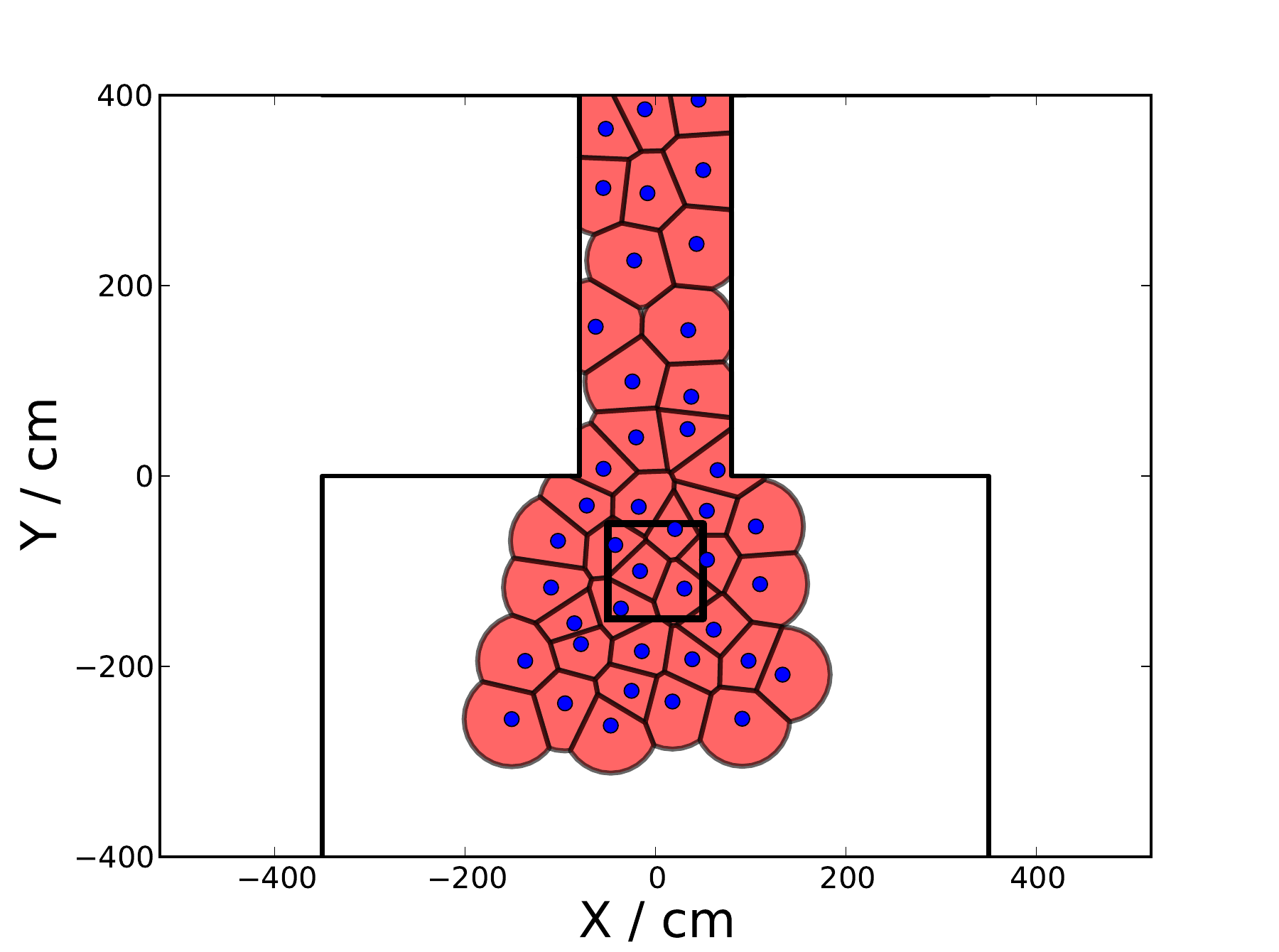}
   \caption{Voronoi diagram.  The cells are all finite, due to \emph{virtual} pedestrian regulation, cutting with geometry and radius regulation.}
   \label{fig:voronoi_diagram}
\end{figure}

Naively calculated Voronoi diagrams possess several features which are inconsistent with the concept of a personal space, in particular the cells on the border of the diagram.  Pedestrians who make up the convex hull have infinite cells, other cells maybe large but finite and many cells cross the boundary of the room.  Three processes are applied to mitigate these problems.  By including \emph{virtual} pedestrians who sit outside the room geometry, all cells are made finite.  The cells are limited to maximum radius (50 cm) and cut with the room geometry.  Figure~\ref{fig:voronoi_diagram} shows a typical diagram after these regulations have been applied.

\section{Application}
We detect non-stationarity in the density for the narrowest bottlenecks (90 cm, 100 cm, 110 cm, 120 cm), which were performed last in the sequence.  This is an important result, as it points to the emergence of a real physical effect, familiarity with the experimental setup does not automatically lead to the development of a stationary-state.  In contrast with~\cite{Kretz2006} where runs were repeated to habituate the participants to the experiment, our test subjects exhibited non-stationary behaviour after having already got used to the experiment.

\begin{figure}
      \centering
      \subfigure[$b = 90$ cm]
      {
         \includegraphics[scale=0.25]{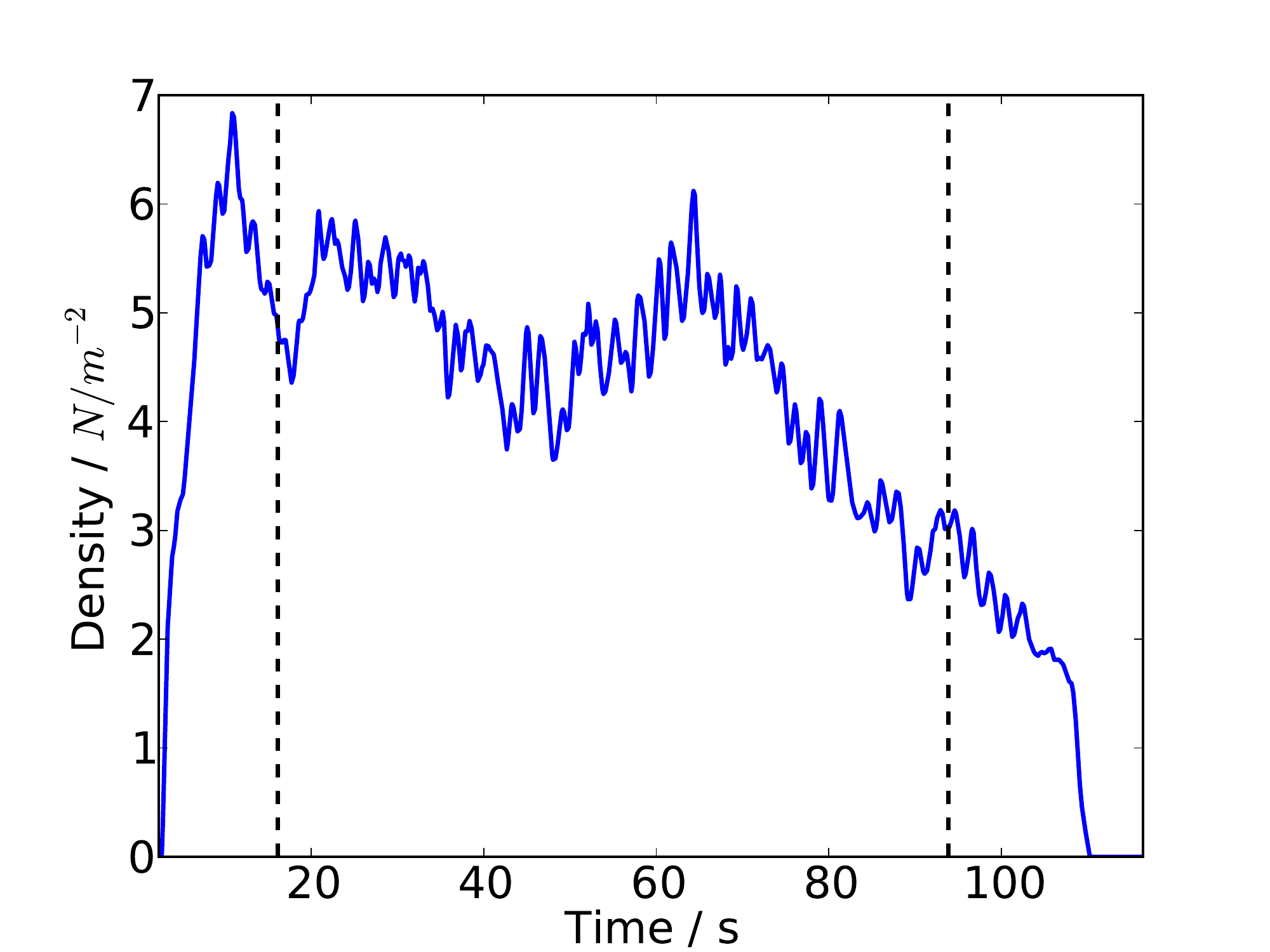}
      }
      \subfigure[$b = 100$ cm]
      {
         \includegraphics[scale=0.25]{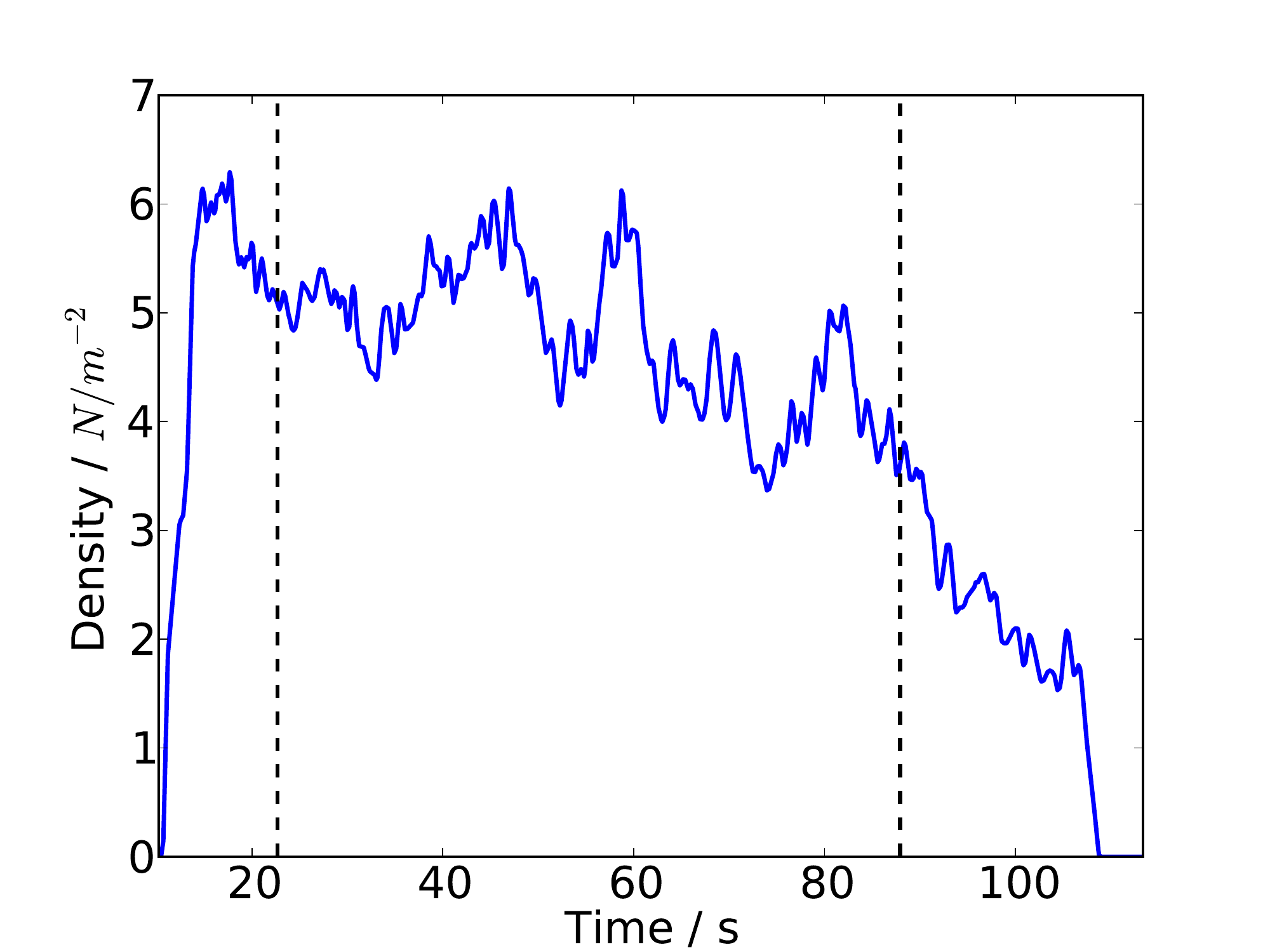}
      }
      \subfigure[$b = 140$ cm]
      {
         \includegraphics[scale=0.25]{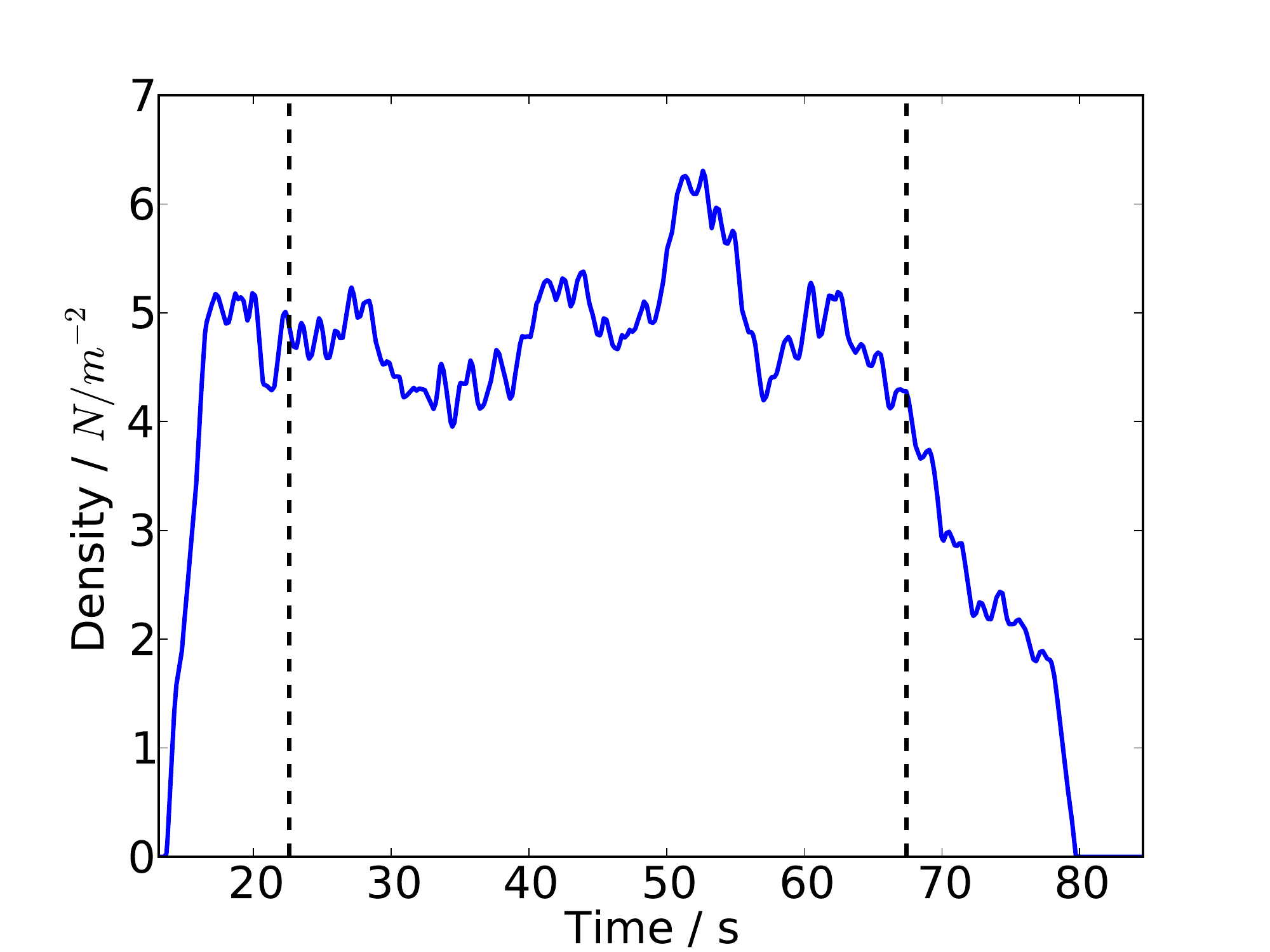}
      }
      \subfigure[$b = 180$ cm]
      {
         \includegraphics[scale=0.25]{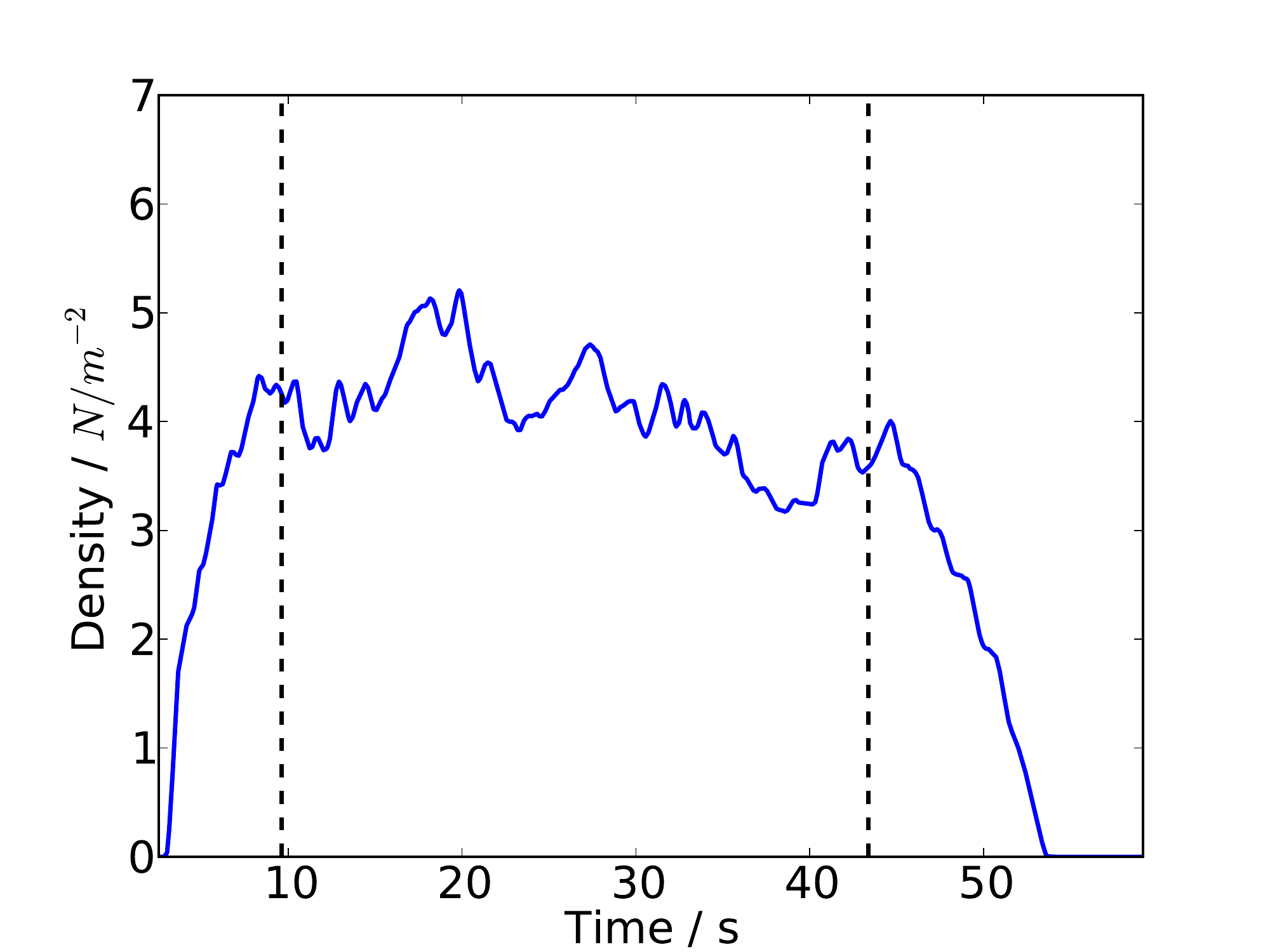}
      }
      \subfigure[$b = 200$ cm]
      {
         \includegraphics[scale=0.25]{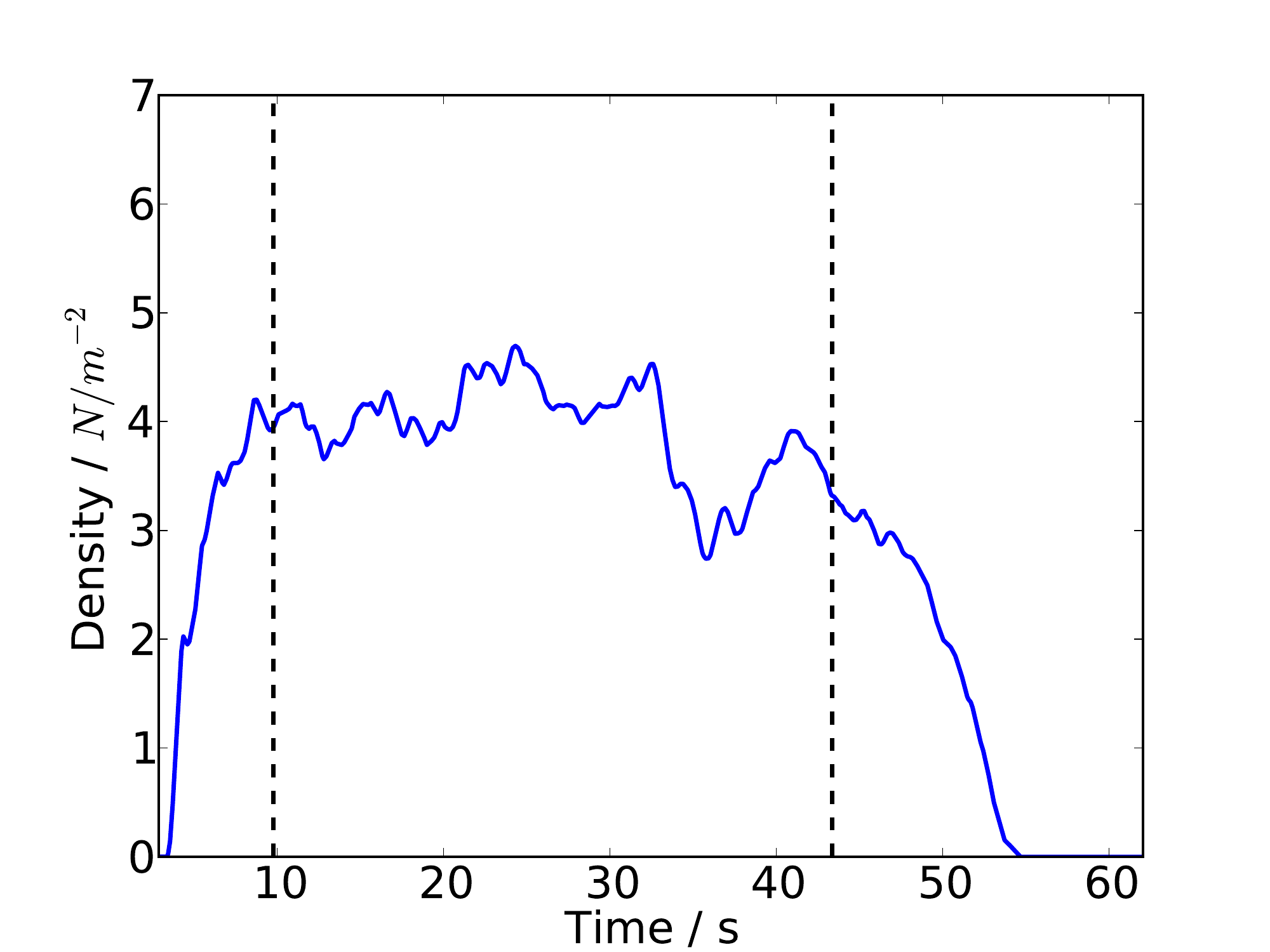}
      }
      \subfigure[$b = 250$ cm]
      {
         \includegraphics[scale=0.25]{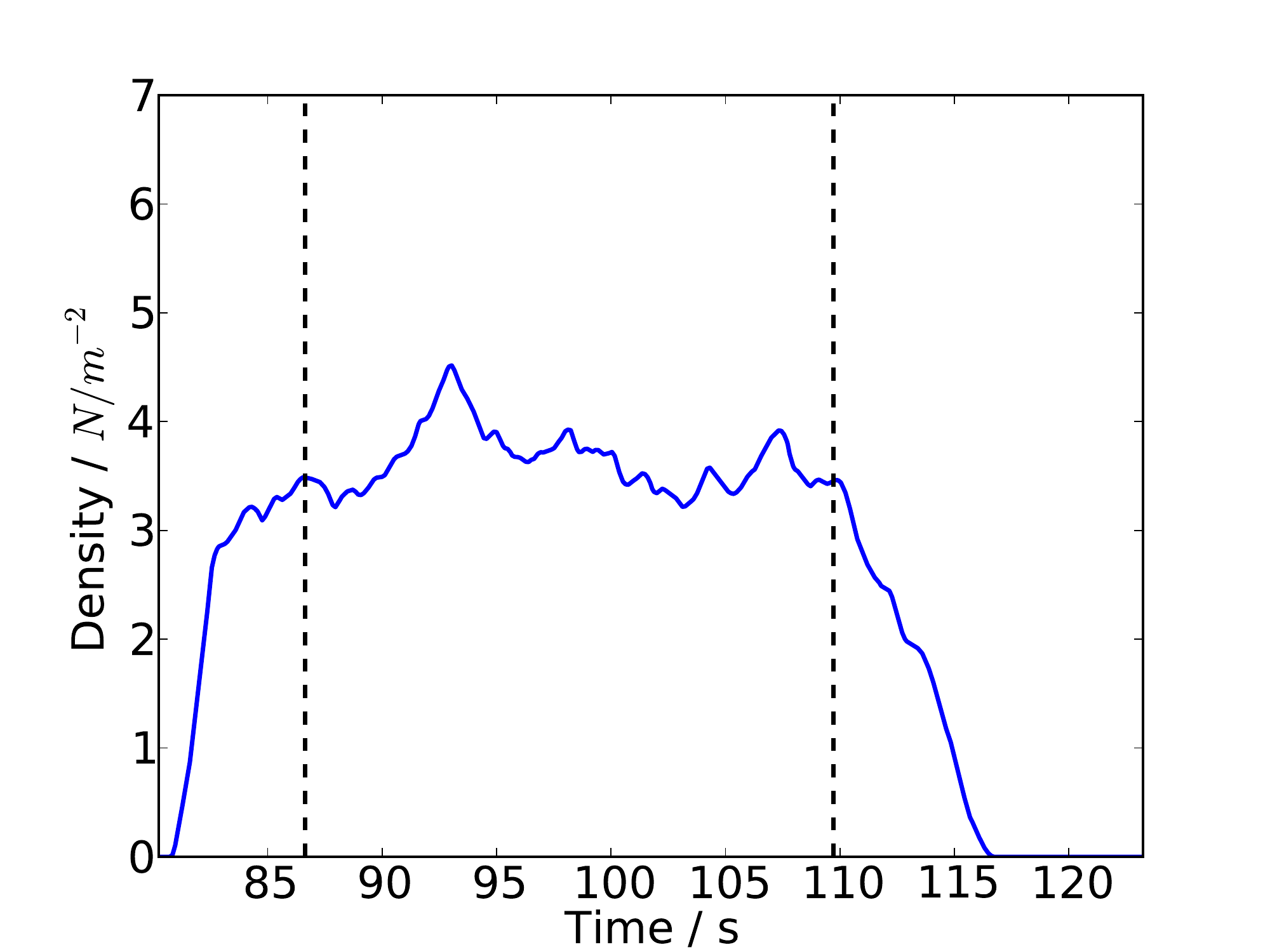}
      }
      \caption{Examples of the integrated density timeseries for different bottleneck widths,$b$.  The location of the cuts is shown.}
      \label{fig:cut_examples}
\end{figure}

At this point in time we are not studying the transient behaviour of the system and this data is removed from consideration.  In order to remove transient behaviour only data after the first and before the last 30 people are considered.  This choice of cuts is chosen from observation of the experiments, after the first 30 have exited, the bottleneck is constantly supplied with pedestrians.  During the egress of the last 30 people the bottleneck is not supplied by a large reservoir of pedestrians.  This choice of cuts is preferred as they can be consistently applied across all experiments.  In figure~\ref{fig:cut_examples} the density time series are shown using our cut criteria.  Note that we are discarding more pedestrians than many experiments used in total.

\clearpage
\subsection{Density, velocity and flow distributions}
\begin{figure}[h]
      \centering
      \subfigure[$b = 140$ cm]
      {
         \includegraphics[scale=0.25]{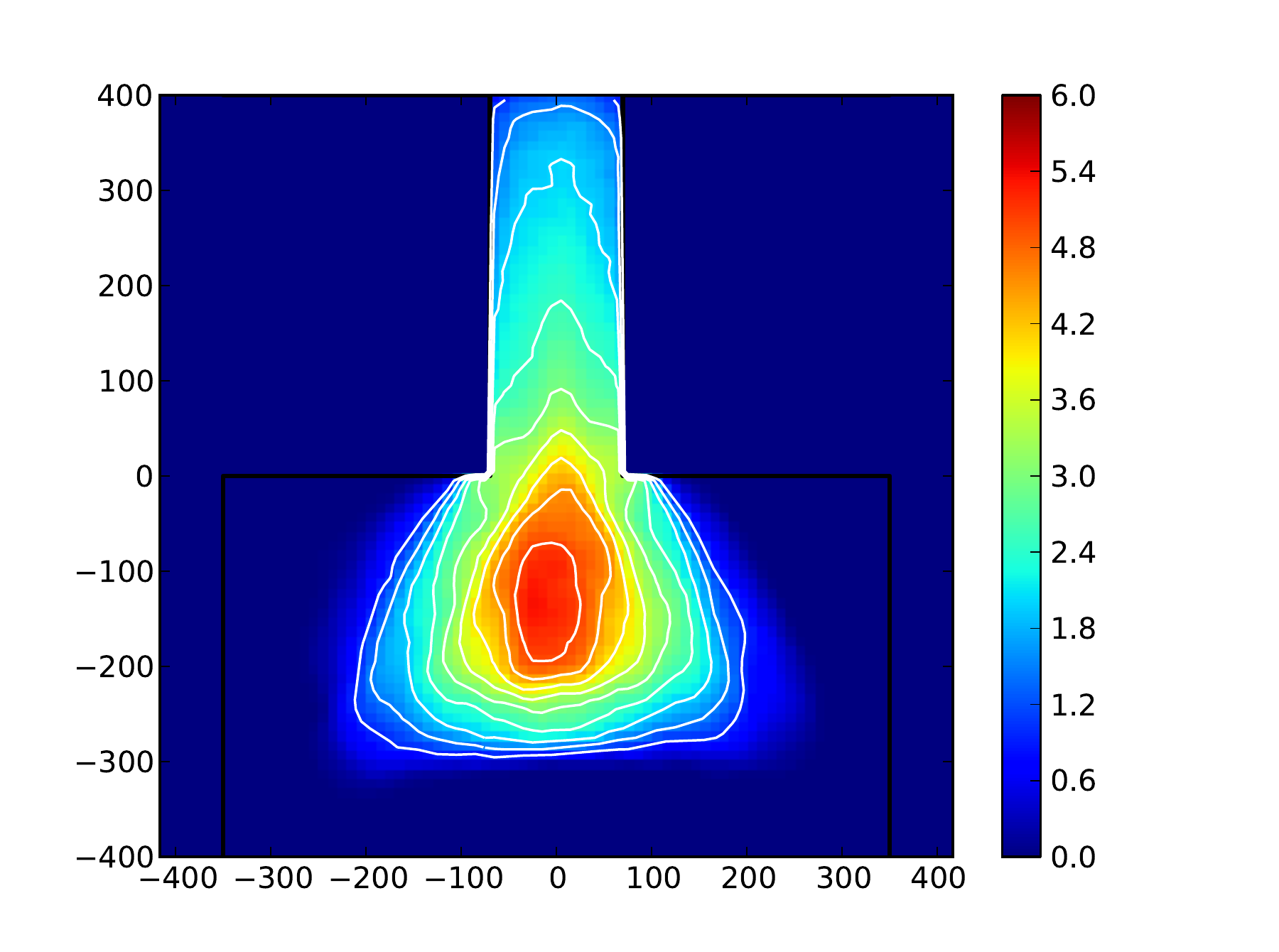}
      }
      \subfigure[$b = 180$ cm]
      {
         \includegraphics[scale=0.25]{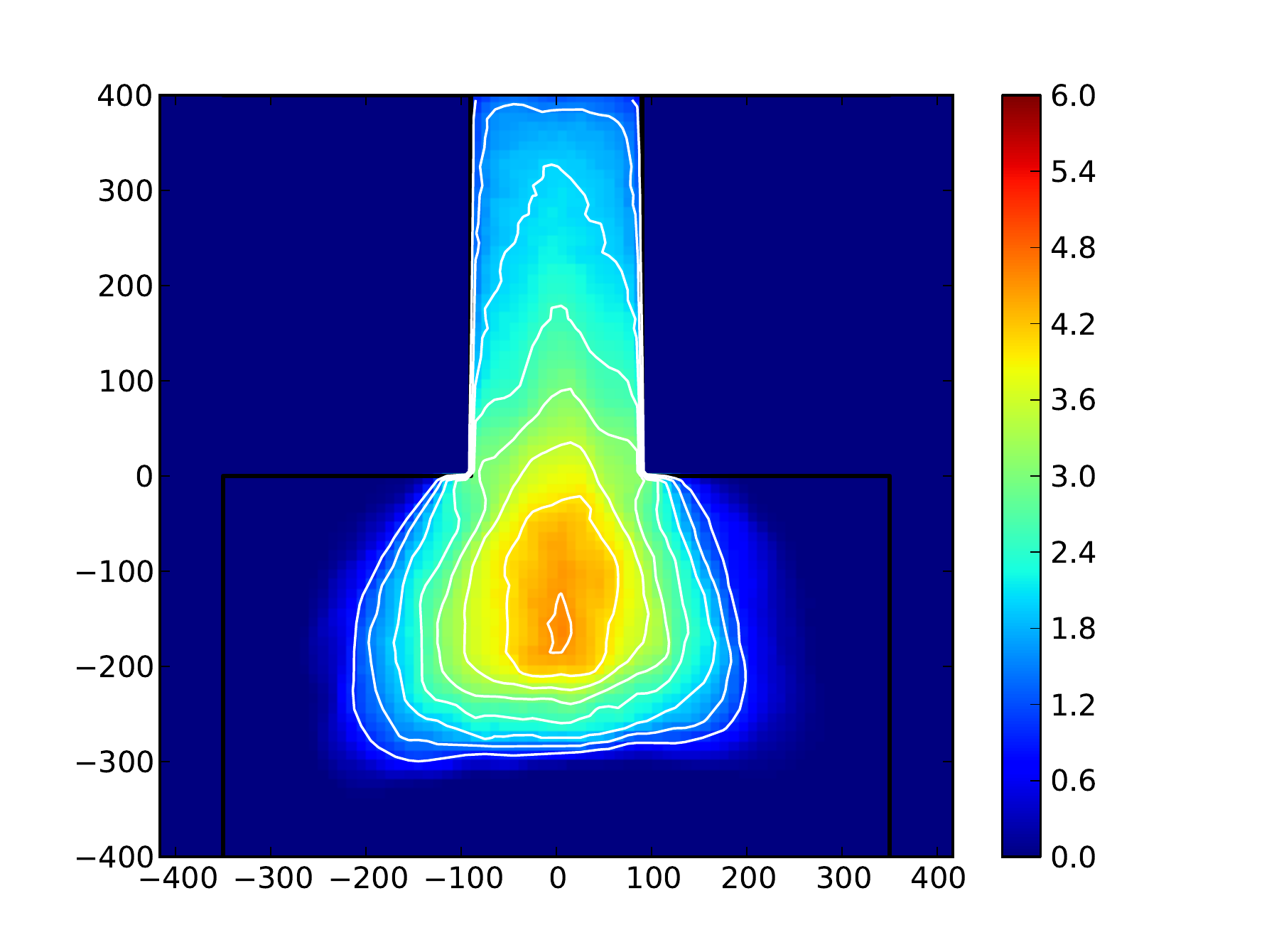}
      }      

      \subfigure[$b = 200$ cm]
      {
         \includegraphics[scale=0.25]{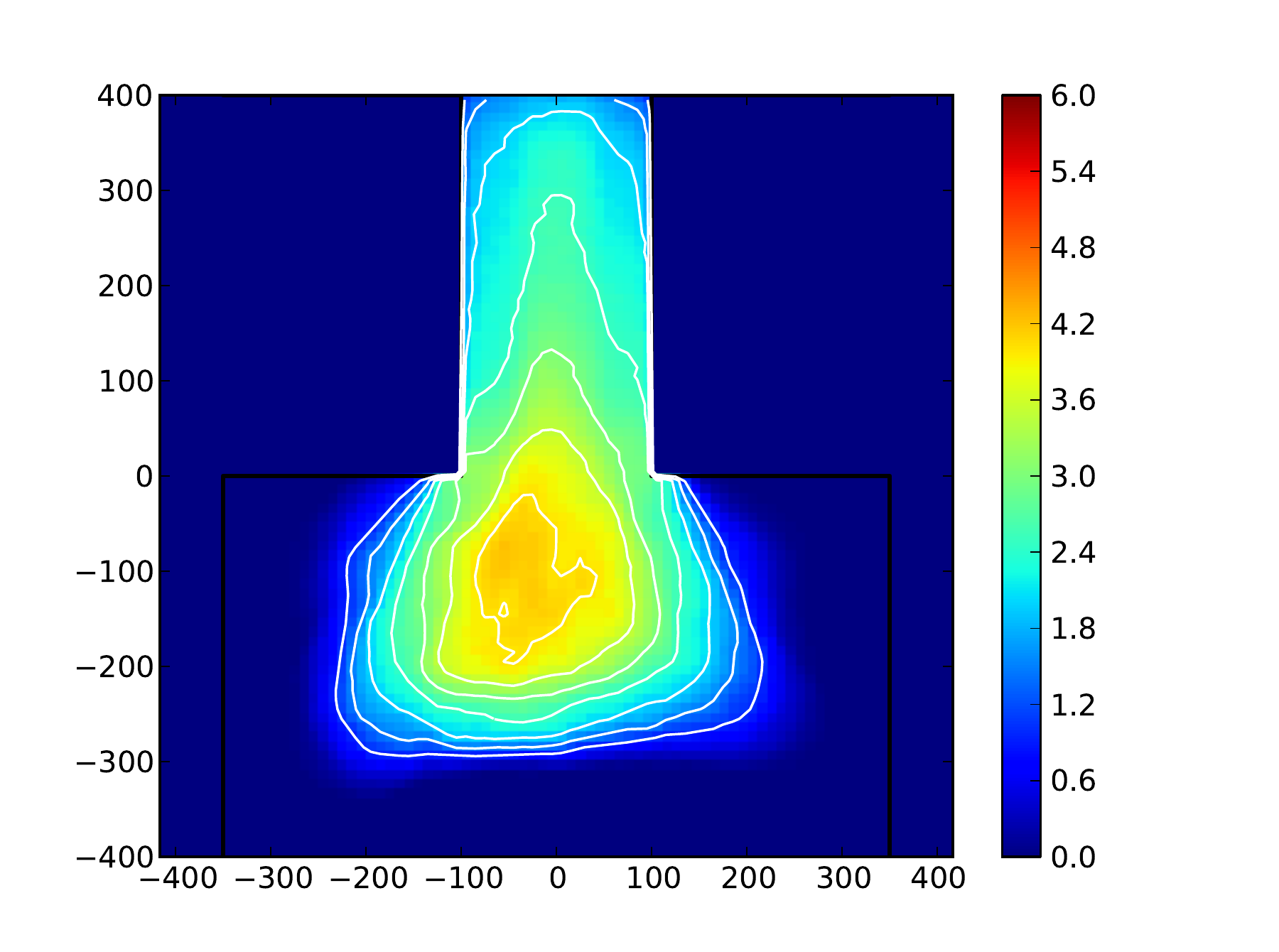}
      }      
      \subfigure[$b = 250$ cm]
      {
         \includegraphics[scale=0.25]{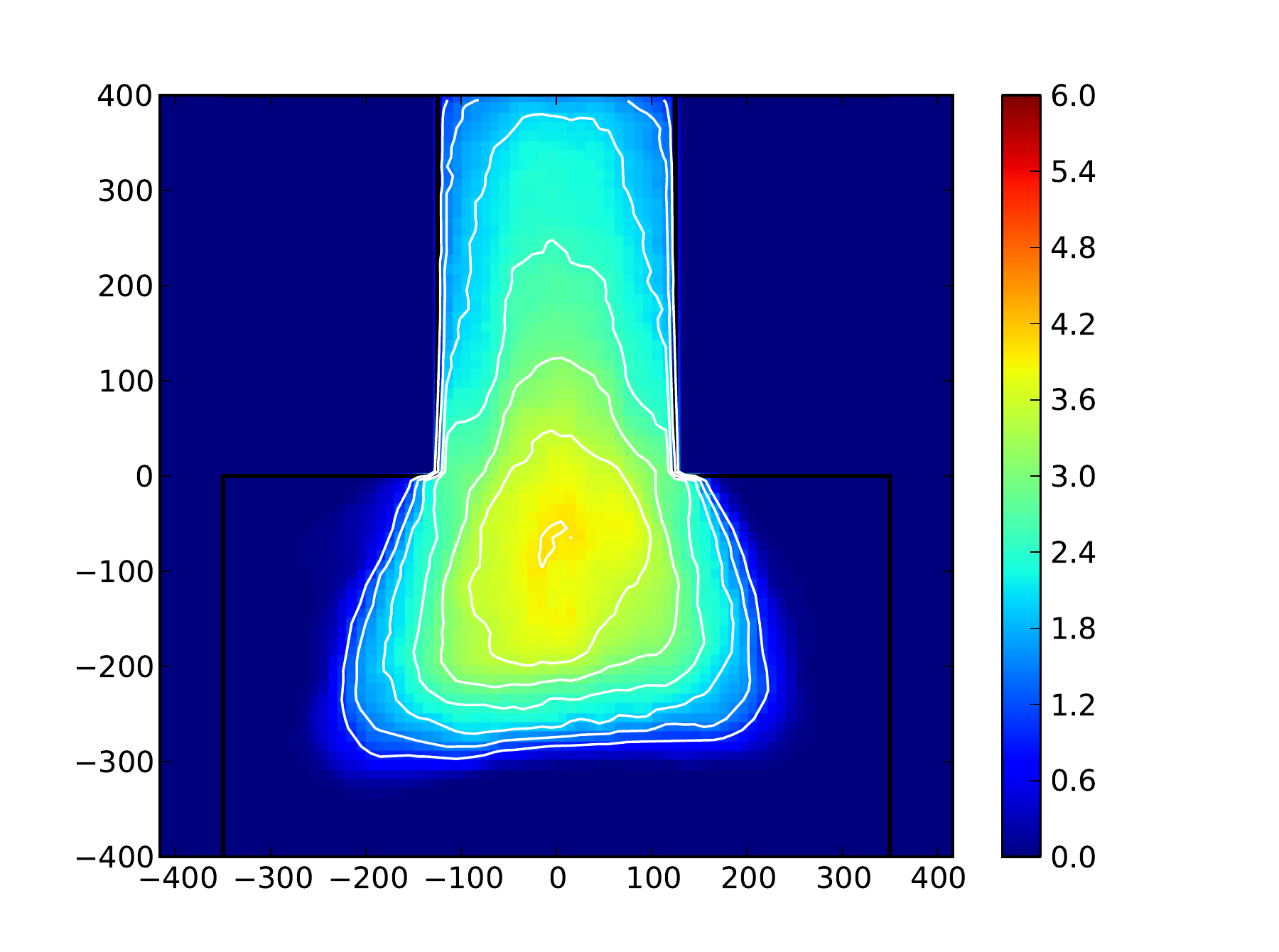}
      }      
      \caption{Integrated density maps for bottlenecks of varying width, $l = 400$ cm.  The contours are placed from $\rho_v = 0.5 $ $P/m^2$ to $\rho_v = 6$ $P/m^2$ in $0.5$ $P/m^2$ intervals.}
      \label{fig:densitymaps_width}      
   \end{figure}

\begin{figure}[h]
      \centering

      \subfigure[$b = 140$ cm]
      {
         \includegraphics[scale=0.25]{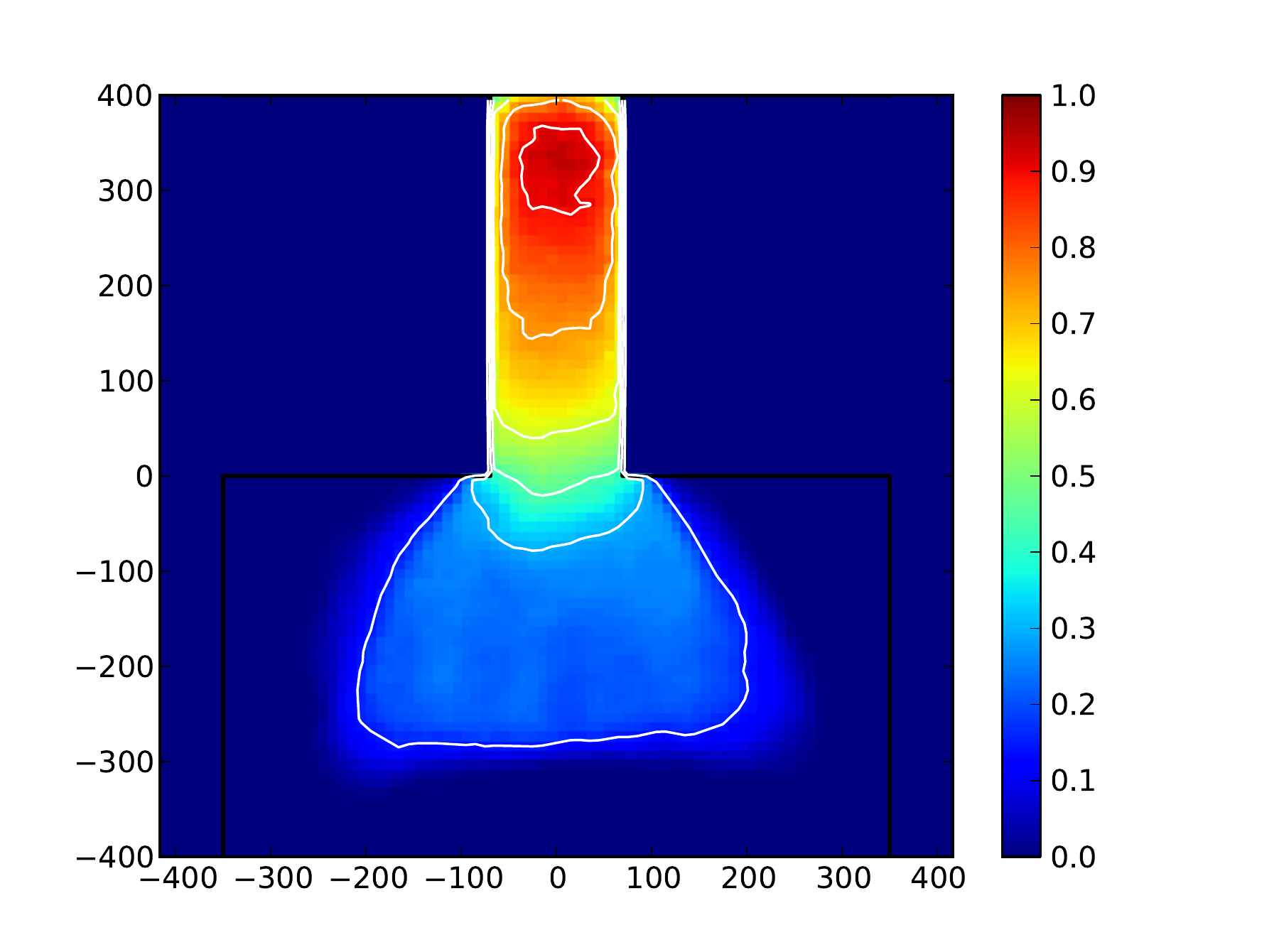}
      }
      \subfigure[$b = 180$ cm]
      {
         \includegraphics[scale=0.25]{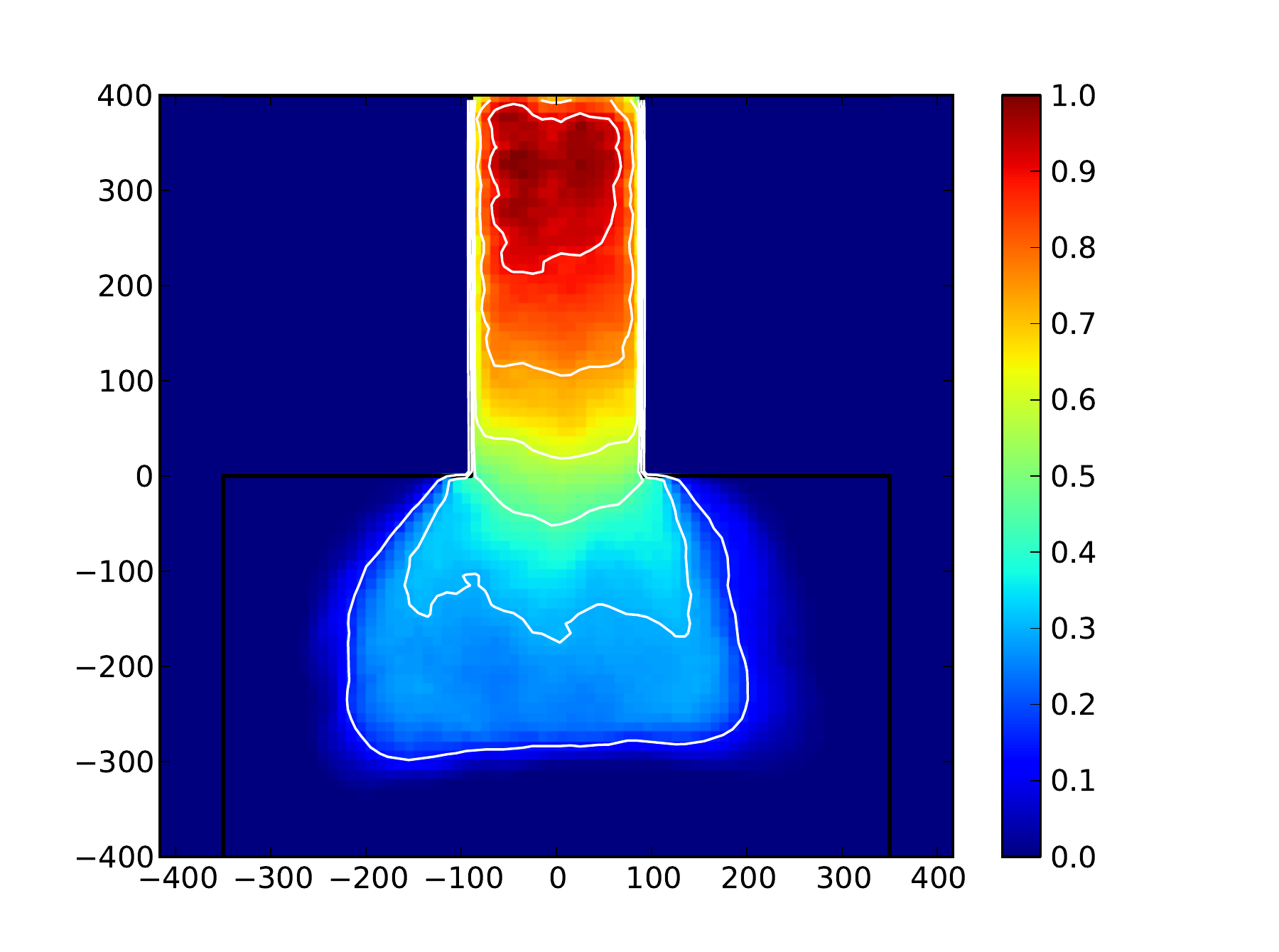}
      }      

      \subfigure[$b = 200$ cm]
      {
         \includegraphics[scale=0.25]{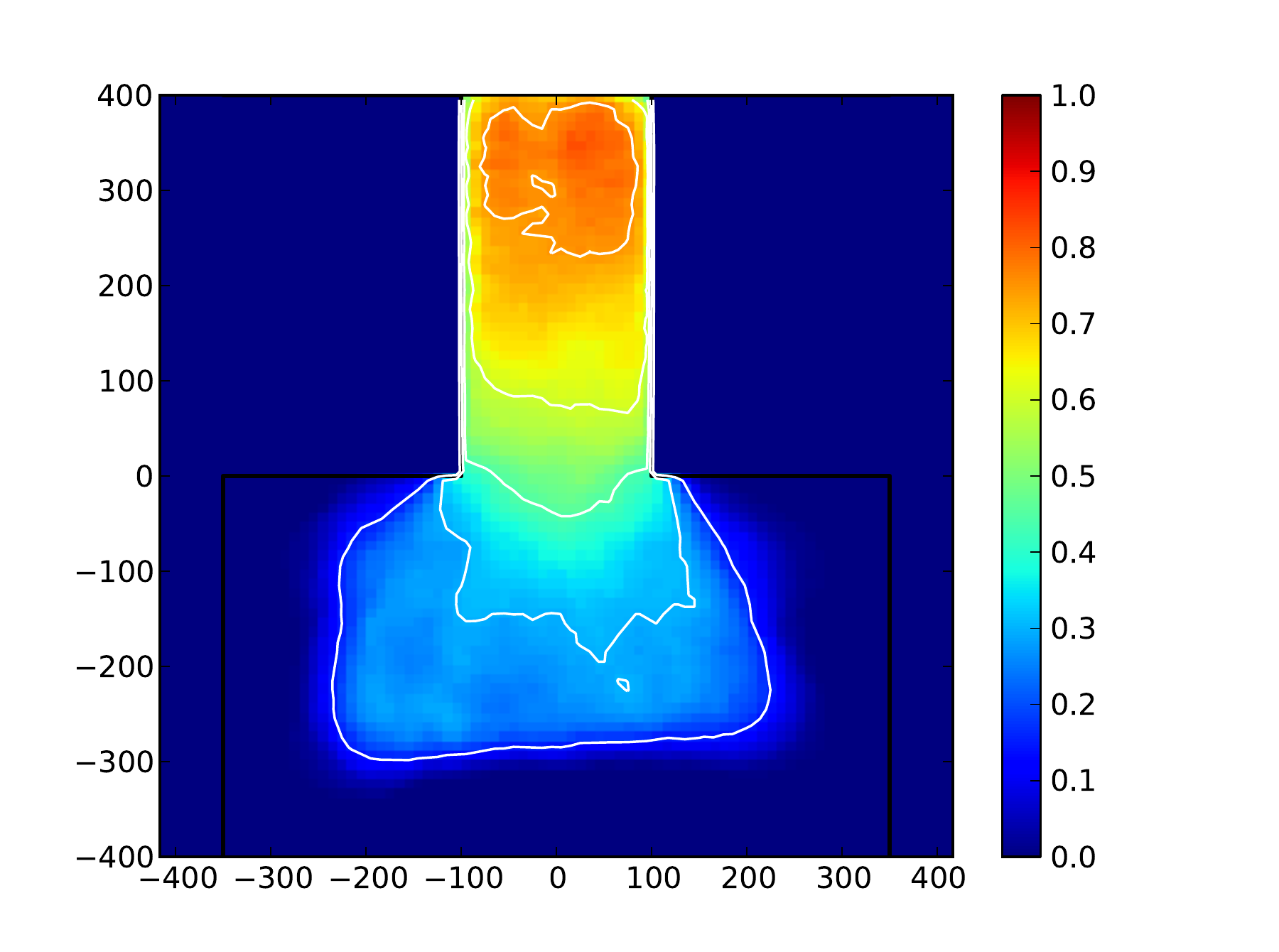}
      }      
      \subfigure[$b = 250$ cm]
      {
         \includegraphics[scale=0.25]{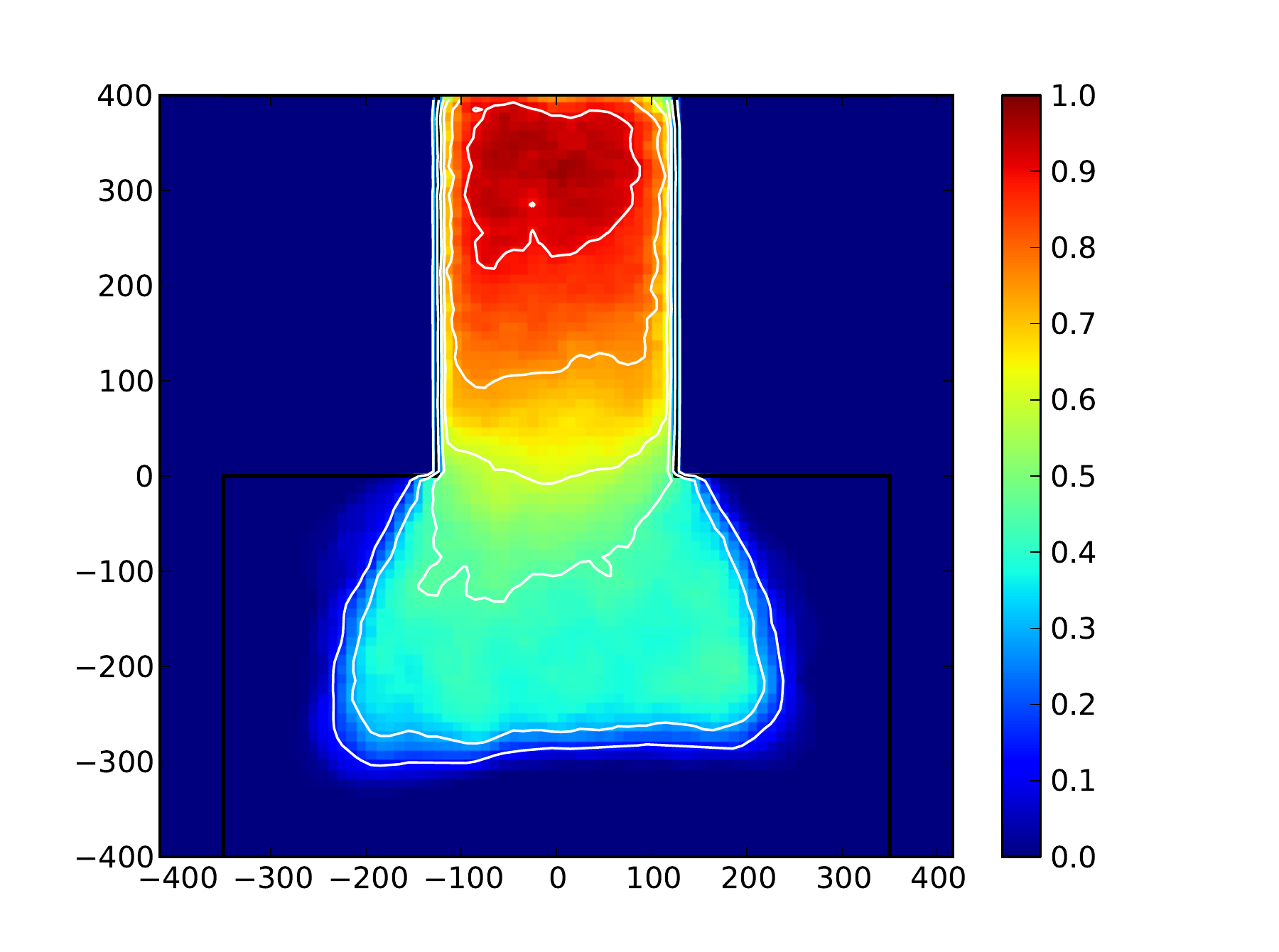}
      }      
      \caption{Integrated velocity maps for bottlenecks of varying width, $l = 400$ cm.}
      \label{fig:velocitymaps_width}      
   \end{figure}

With the integrated density method the measurement area may be smaller than the pedestrians, something which cannot be done with standard measurements.  By considering the integrated density, velocity and flow calculated over small (10 cm $\times$ 10 cm) regions and averaging over all stationary state times, maps of the observables over the experimental area can be obtained, figures~\ref{fig:densitymaps_width} and~\ref{fig:velocitymaps_width}.  These maps are presented with contours to present additional information, such as the location of the maximum density and directions of steepest decline.  These maps provide insights into the dynamics of the motion and the sensitivity of the integrated quantities to influences such as boundary geometry and measurement placement.  We see that the integrated density is sensitive to the geometry, as the Voronoi cells are regulated by the geometry, and the cutoff radius applied to the cells.  We only present maps for bottlenecks which exhibited a stationary state.

Immediately observable is the strength of the peak in front of the bottleneck.  Narrower bottlenecks have a higher density in front of the constriction and the peak in the density is sharper.  The peak in the density extends further into the bottleneck for the wider bottlenecks studied.

\begin{figure}[h]
      \centering

      \subfigure[$b = 140$ cm]
      {
         \includegraphics[scale=0.25]{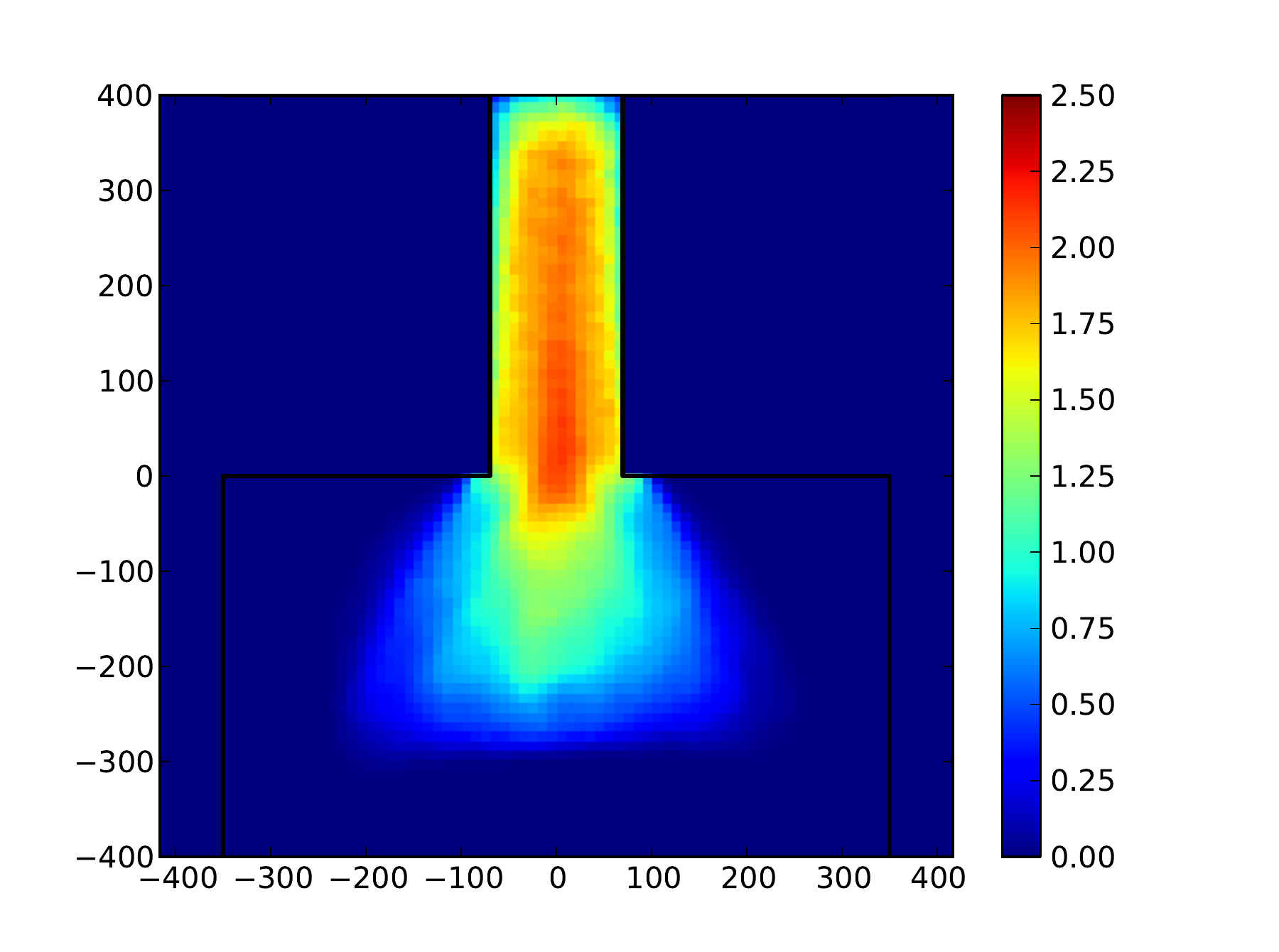}
      }
      \subfigure[$b = 180$ cm]
      {
         \includegraphics[scale=0.25]{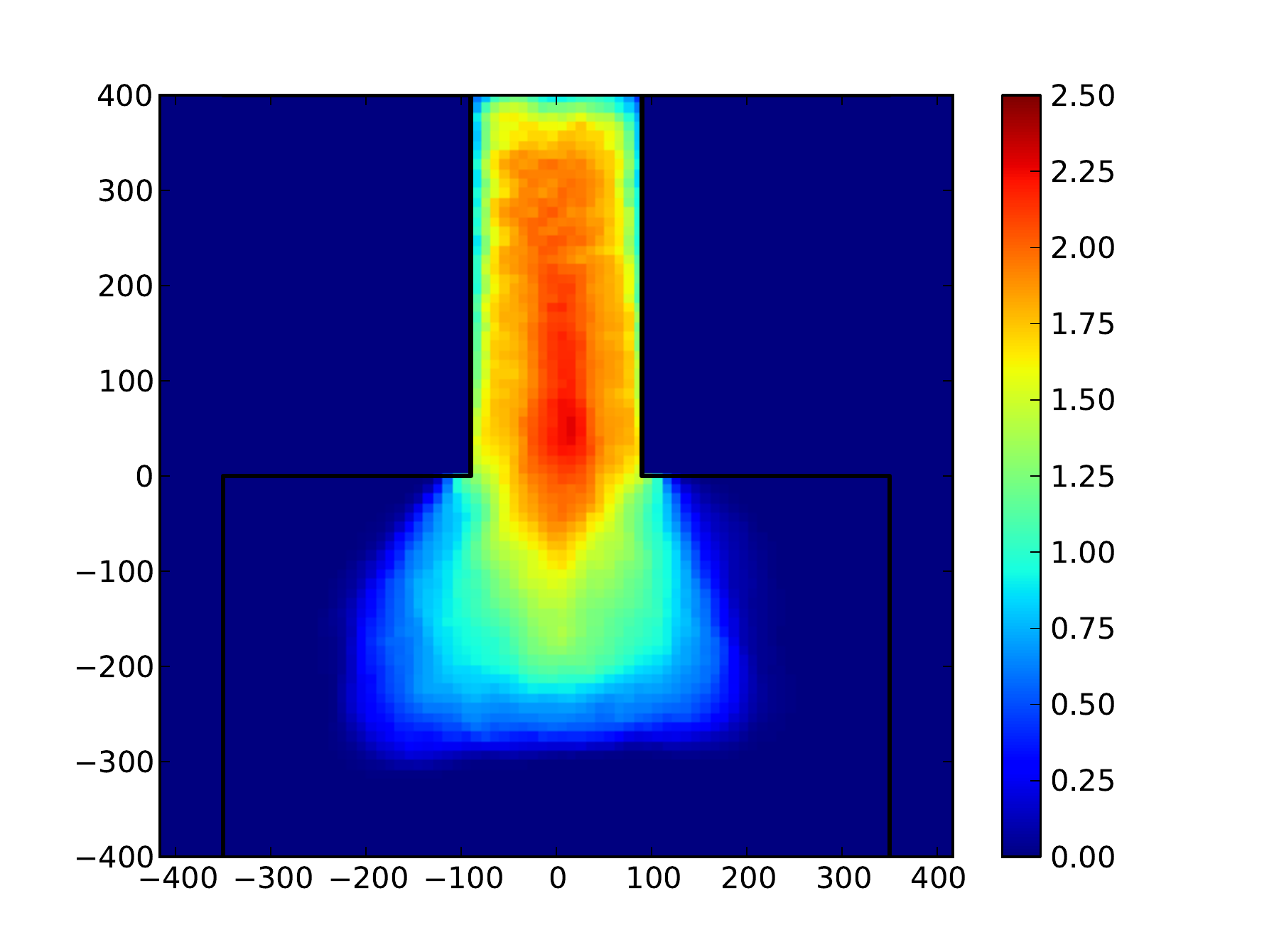}
      }      

      \subfigure[$b = 200$ cm]
      {
         \includegraphics[scale=0.25]{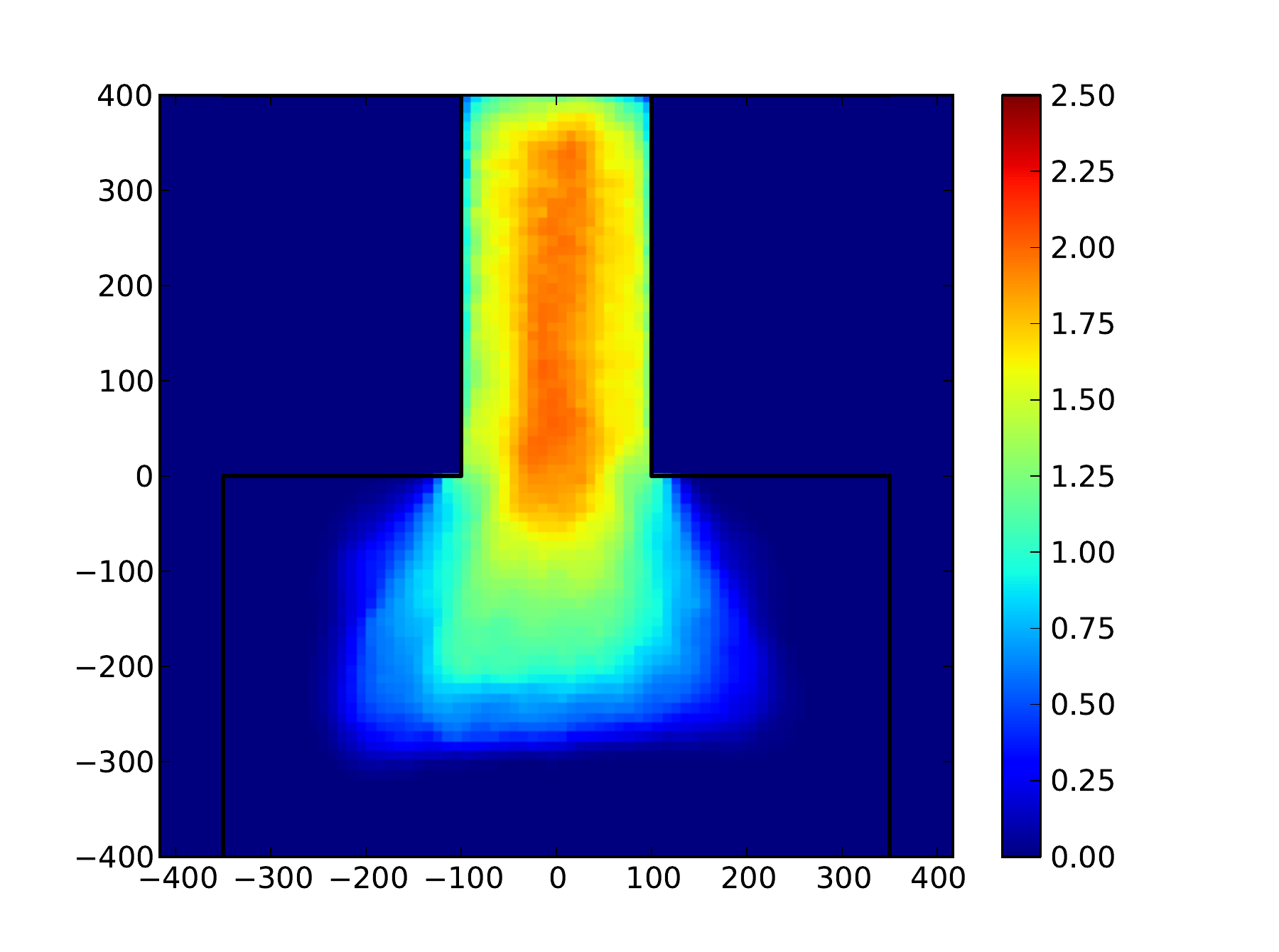}
      }      
      \subfigure[$b = 250$ cm]
      {
         \includegraphics[scale=0.25]{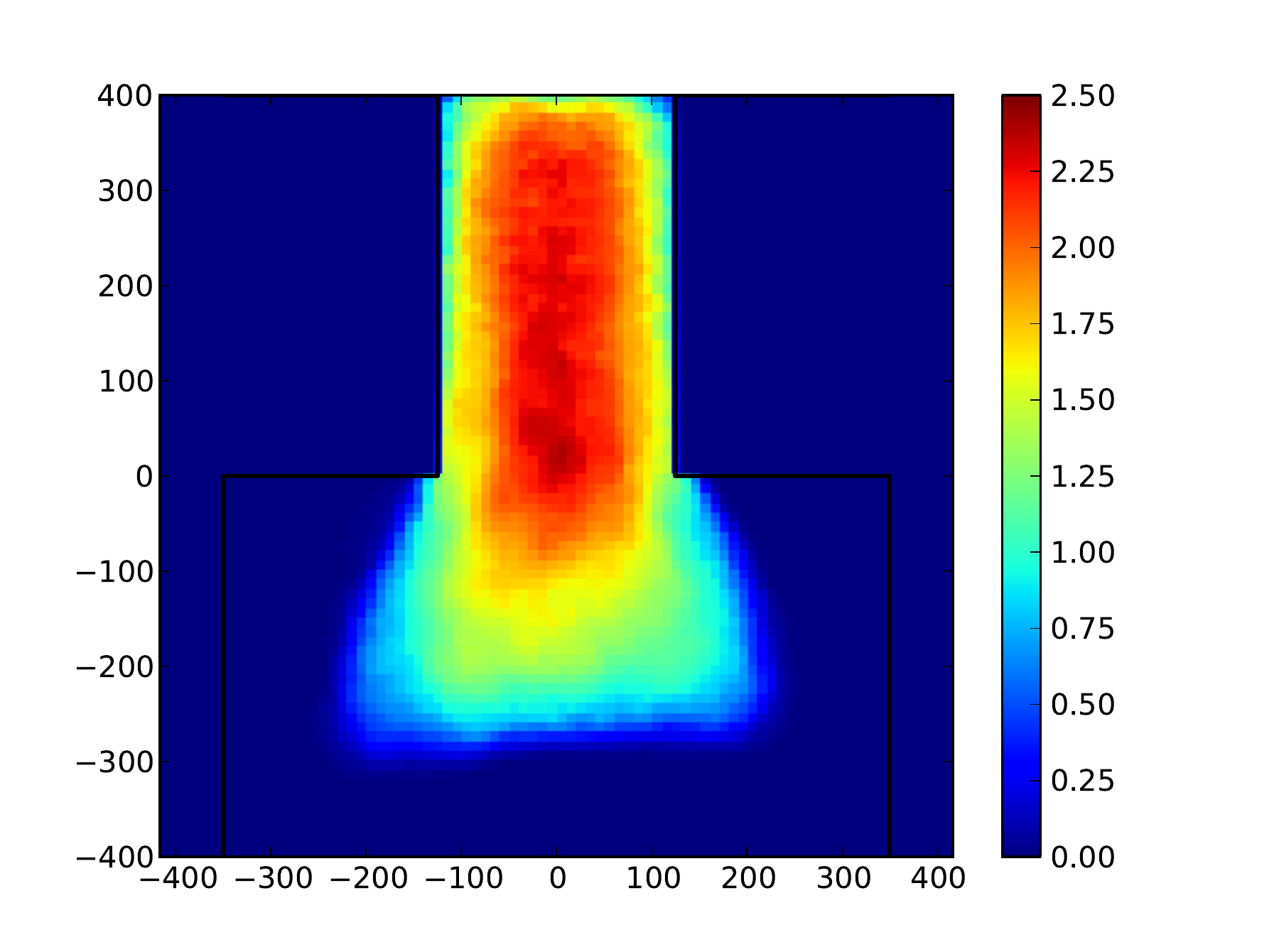}
      }      
      \caption{Specific flow, $J_s$. for bottlenecks of varying width, $l = 400$ cm.}
      \label{fig:flowmaps_width}      
   \end{figure}

In figure~\ref{fig:flowmaps_width} the hydrodynamic flow is shown.  The highest flow is seen in the channel and the high flow region extends further into the waiting area as the bottleneck width increases.

\clearpage

Density and velocity measurements are sensitive to the size and location of the measurement area.  This is unavoidable, as we are approximating a local parameter, e.g. $\rho(x)$, with a measurement over a finite area.  Due to this, no measure can be considered more correct than any other.  When considering non-local approximations the dependence on the geometry and location of the measurement area must be considered.  From the maps it is clear that the size and location of the measurement area will influence the measured density.


\subsection{Choice of measurement area}
The integrated density follows the same trends as the standard density, however we should investigate whether or not any systematic difference can be seen.  To do this the measurement area, $A$, must be placed and dimensioned with care.  Three different effects should be considered in choosing a measurement area: the location of the measurement area, the shape of the measurement area and the choice of measurement method.  The criteria for choosing the location of the measurement is simple, we wish to choose an area where the sensitivity to measurement location and size is minimized.  Due to the symmetry of the experimental setup the observation area is chosen to lie directly in front of the constriction, symmetrically about the middle.

We wish to capture information about all the pedestrians passing through the bottleneck.  As the bottleneck widens the peak of the density spreads in the horizontal direction.  This implies that we should widen the observation patch with the bottleneck.  The measurement area is chosen to be 20 cm narrower than the bottleneck, and to be 100 cm deep, approximately the depth of two pedestrians.

We wish to place the centre of the measurement area at or near to the peak in the density, it is at this point where the measurement will be least sensitive to changes in the measurement location. The density maps show that this peak lies about 125 cm from the bottleneck entrance.  The peak gets a little closer (around 100 cm) for the widest bottlenecks.

\begin{figure}
   \centering
   \includegraphics[scale=0.5]{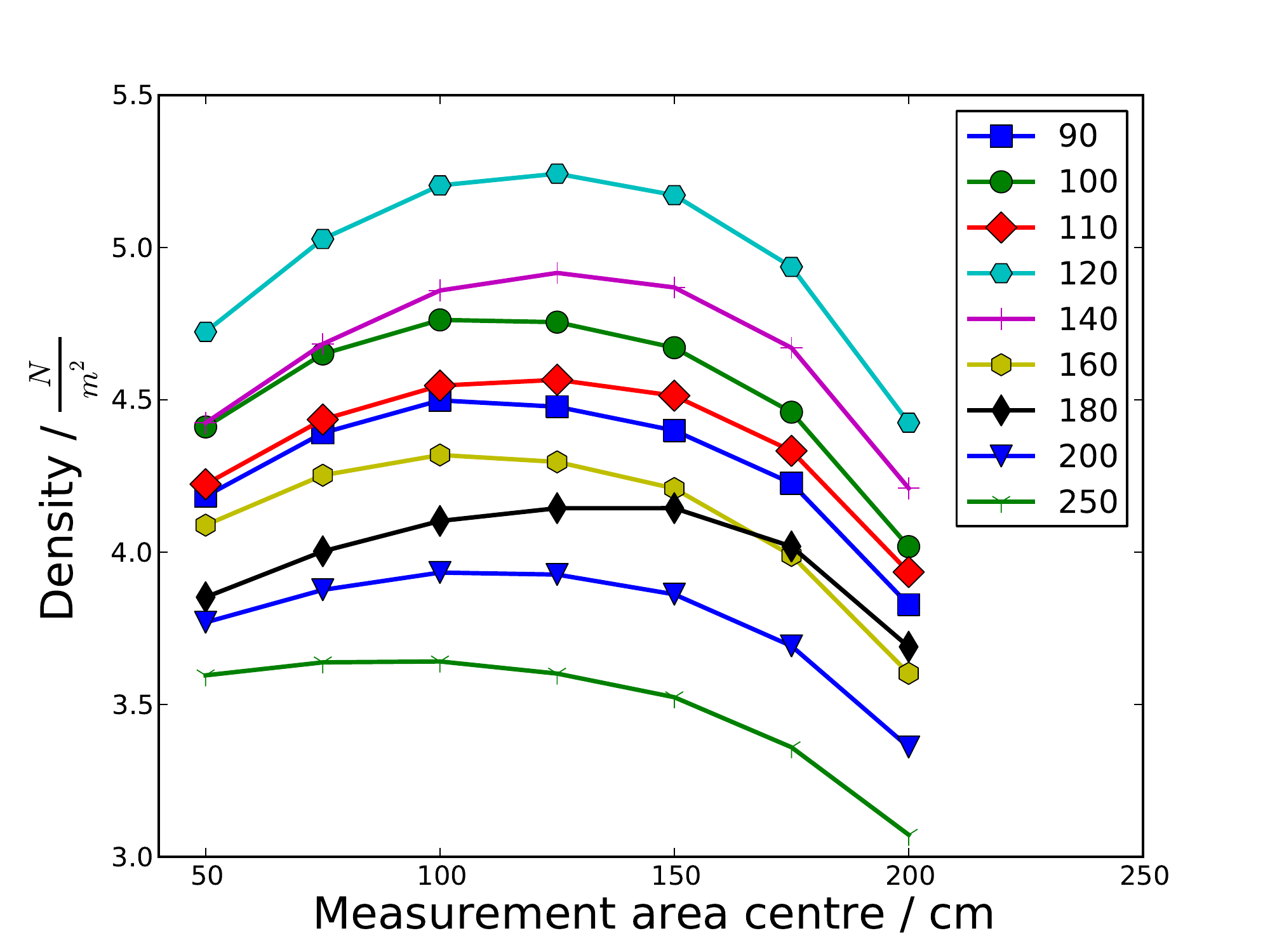}
   \caption{Change in integrated density with location of measurement area.  The peak at around 125 cm is clear visible.}
   \label{fig:measurement_location}
\end{figure}

In figure~\ref{fig:measurement_location} the sensitivity of the density to the location of the measurement area is shown.  The trend of the peak maximum towards the entrance as the bottleck widens can be seen.

\begin{figure}
   \centering
   \includegraphics[scale=0.5]{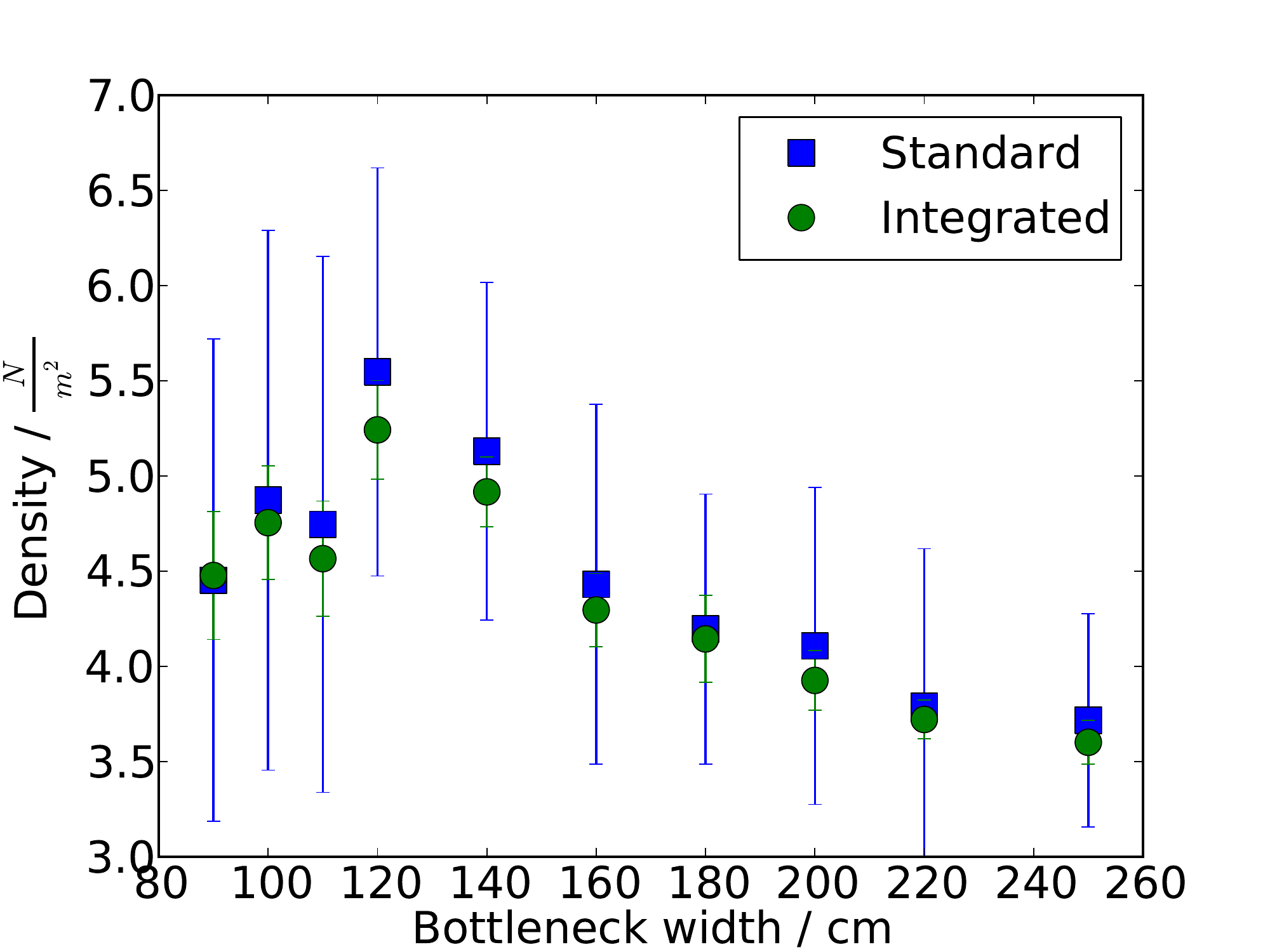}
   \caption{Discrepancy between the integrated and standard densities.  The error bars represent our estimate of the size of the fluctuations.  The low density values seen with the narrowest bottlenecks are attributed to the non-stationarity seen in these experiments.}
   \label{fig:method_changing1}
\end{figure}

In figure~\ref{fig:method_changing1} the difference between the two measurement methods can be seen, using our rectangular measurement area.  The integrated density is consistently lower than the standard density (although well within fluctuations).  An explanation lies in the scale sensitivity of the two methods.  The integrated method is sensitive on scales larger than the measurement area.  A pedestrian passing near the measurement area will contribute to the integrated density (contributing a relatively low density as it sits away from the maximum) but make no contribution to the classical density.

\section{Flow through bottleneck}

\begin{figure}[ht]
   \centering   
   \includegraphics[scale=0.5]{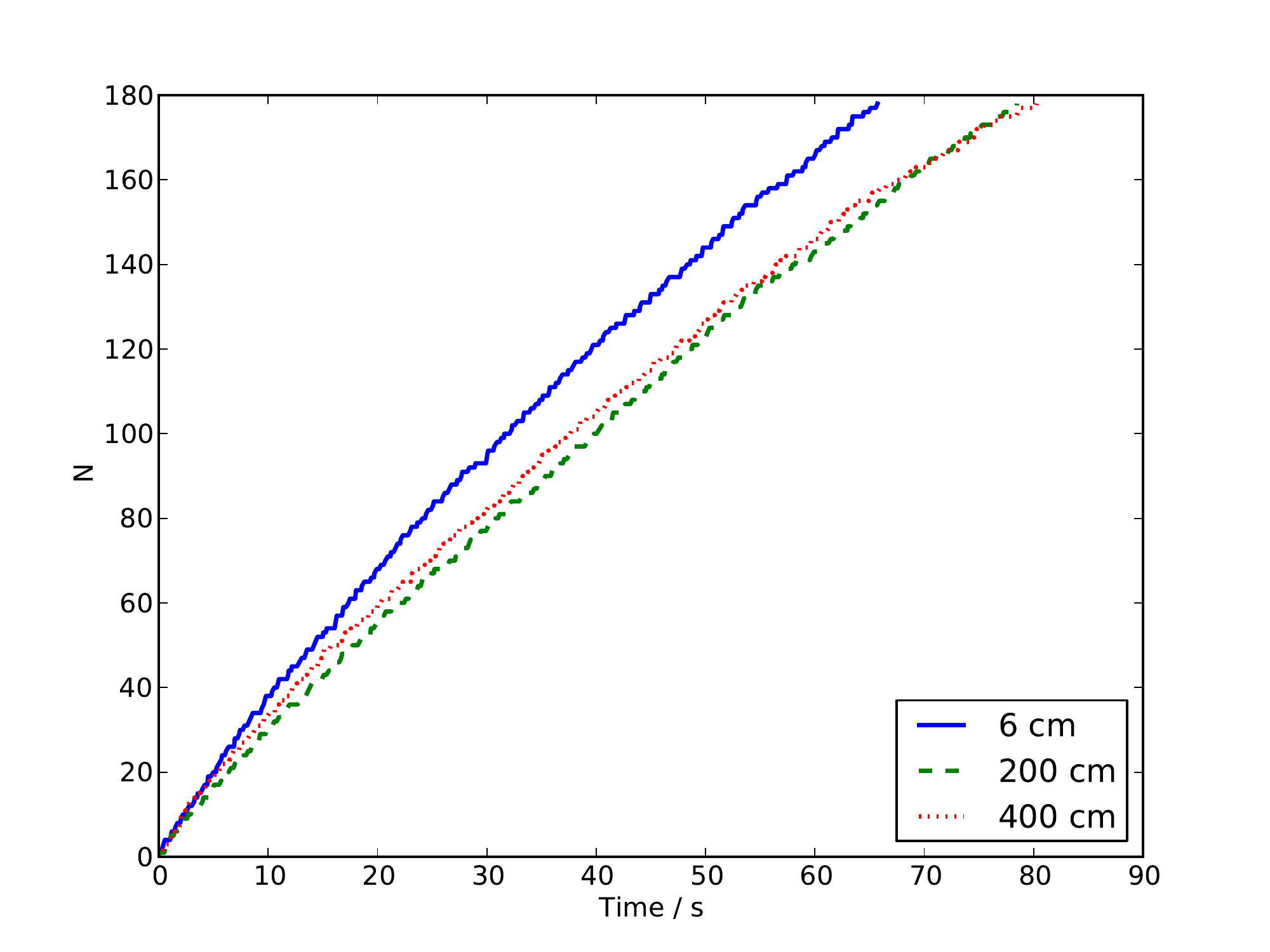}
   \caption{$N(t)$, the number of people having passed the entrance of the bottleneck.  For bottlenecks of varying length and equal width (b = 120 cm).}
   \label{fig:length_N}
\end{figure}

In~\cite{Liddle2009} it was observed that the flow through the shortest bottlenecks was noticeably greater than the flow through the longer bottlenecks.  This can be seen in figure~\ref{fig:length_N} in a given time period more pedestrians egress through the shorter bottlenecks than the longer bottlenecks.

\begin{figure}[ht]
   \centering
   \includegraphics[scale=0.5]{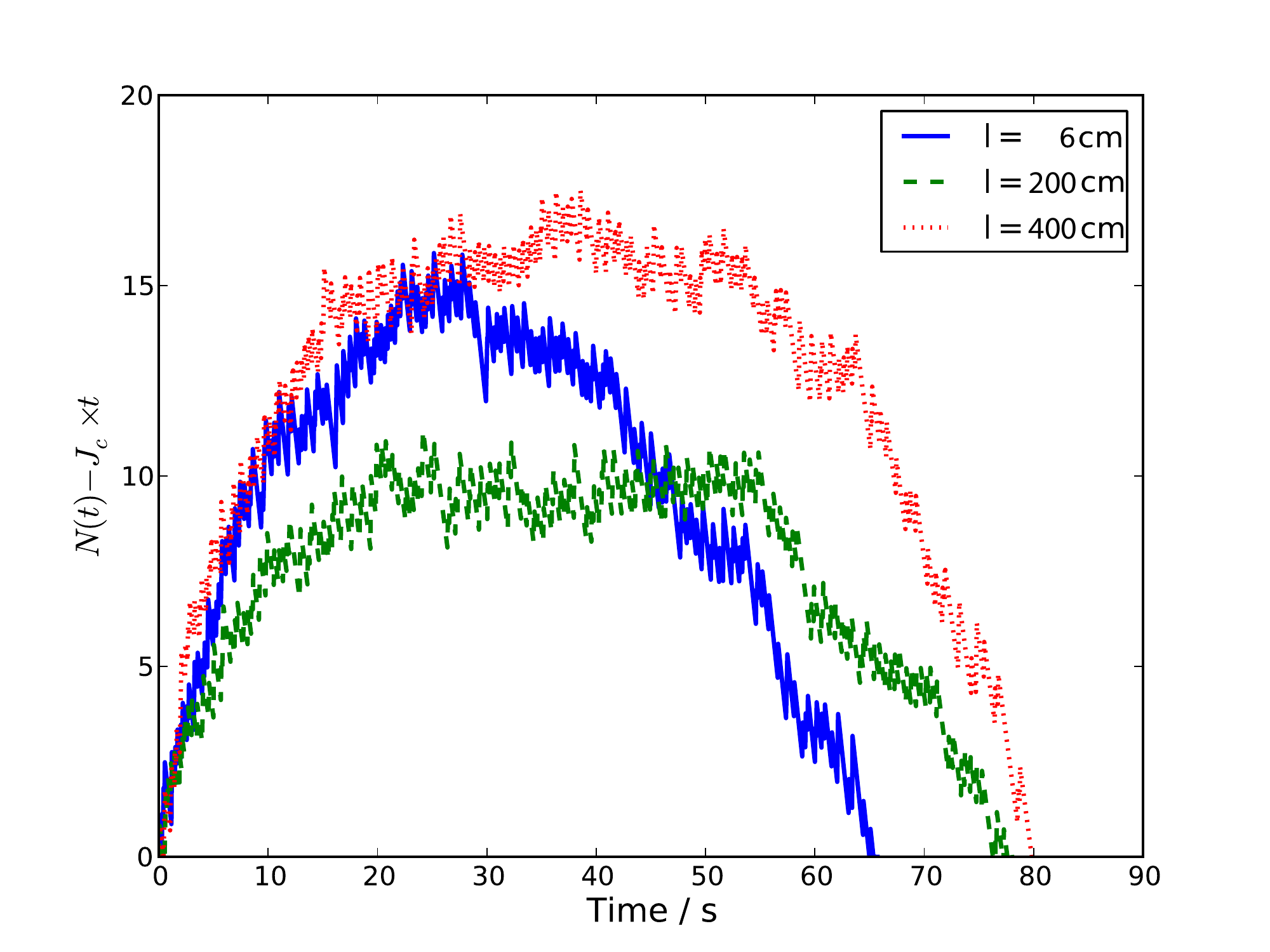}
   \caption{$N(t)-J_c \times t$ for, the number of people having passed through the bottleneck minus a constant flow contribution.  The flat and linear areas correspond to regions of stationary flow.}
   \label{fig:length_deltaN}
\end{figure}

To inspect the time dependence in a clearer manner we subtract the observed pedestrian count $N$ from $J_c \times t$, where $J_c$ is the average flow over the entire experiment.  In figure~\ref{fig:length_deltaN} this is plotted.  Straight sections correspond to periods of constant flow, positive gradients.  For $l = 200$ cm, $400$ cm the horizontal sections between $t = 20$ s and $t= 45$ s show us that during this period the flow is constant with a value equal to $J_c(l)$.  However the $l = 6$ cm runs exhibit no regions of constant flow.

The integrated densities ability to probe microscopic scales can be applied here to reveal the underlying physical cause.

In figure~\ref{fig:rhov_l} the magnitude of the specific flow,
\begin{equation}
   |J_s| = \rho |v|,
\end{equation}
is shown for the three varying length experiments performed (calculated over 10 cm square regions).  As is expected for the long bottlenecks the specific flow is constant inside the bottleneck (figures~\ref{fig:rhov_l_200}, \ref{fig:rhov_l_400}).  In all the experiments a protrusion out into the waiting area can be seen.  The peak in the specific flow is highest for the shortest bottleneck.  These peaks can be seen clearer in figure~\ref{fig:rhov_slice}.

\begin{figure}[ht]
   \centering
   \subfigure[$l = 6$ cm]
   {
      \includegraphics[scale=0.4]{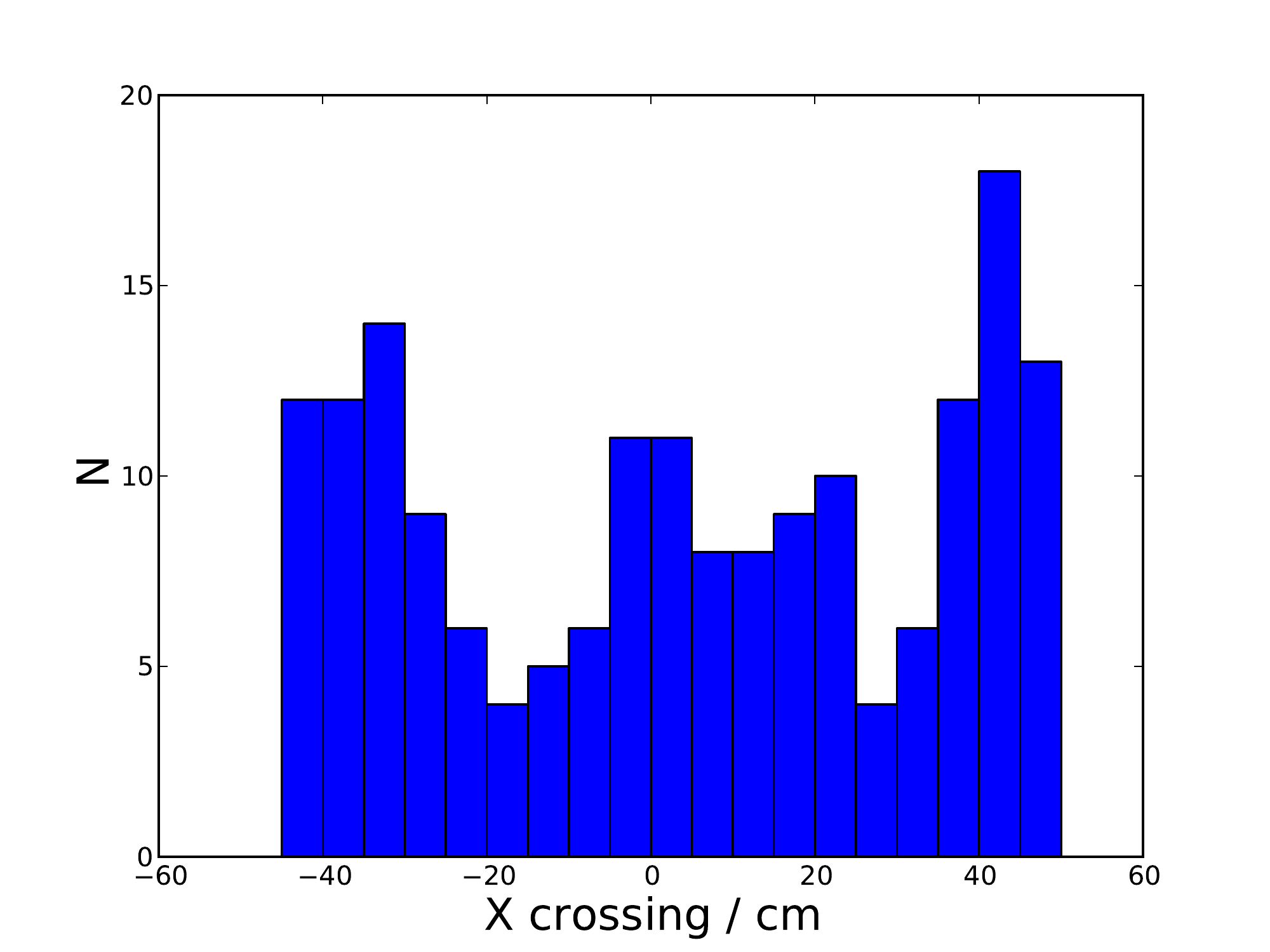}
      \label{fig:rhov_l_006}
   }
   \subfigure[$l = 200$ cm]
   {
      \includegraphics[scale=0.4]{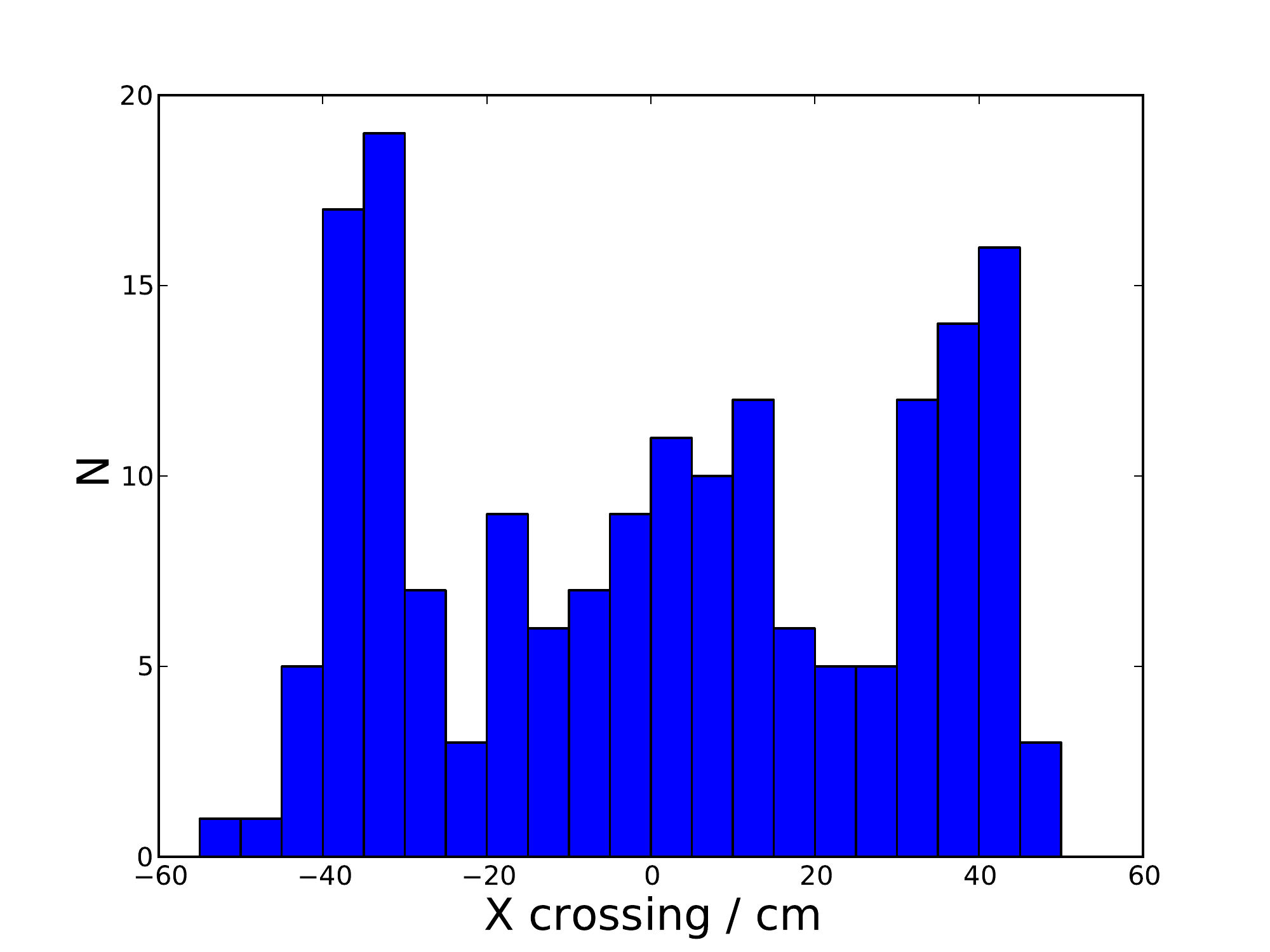}
      \label{fig:rhov_l_200}
   }
   \subfigure[$l = 400$ cm]
   {
      \includegraphics[scale=0.4]{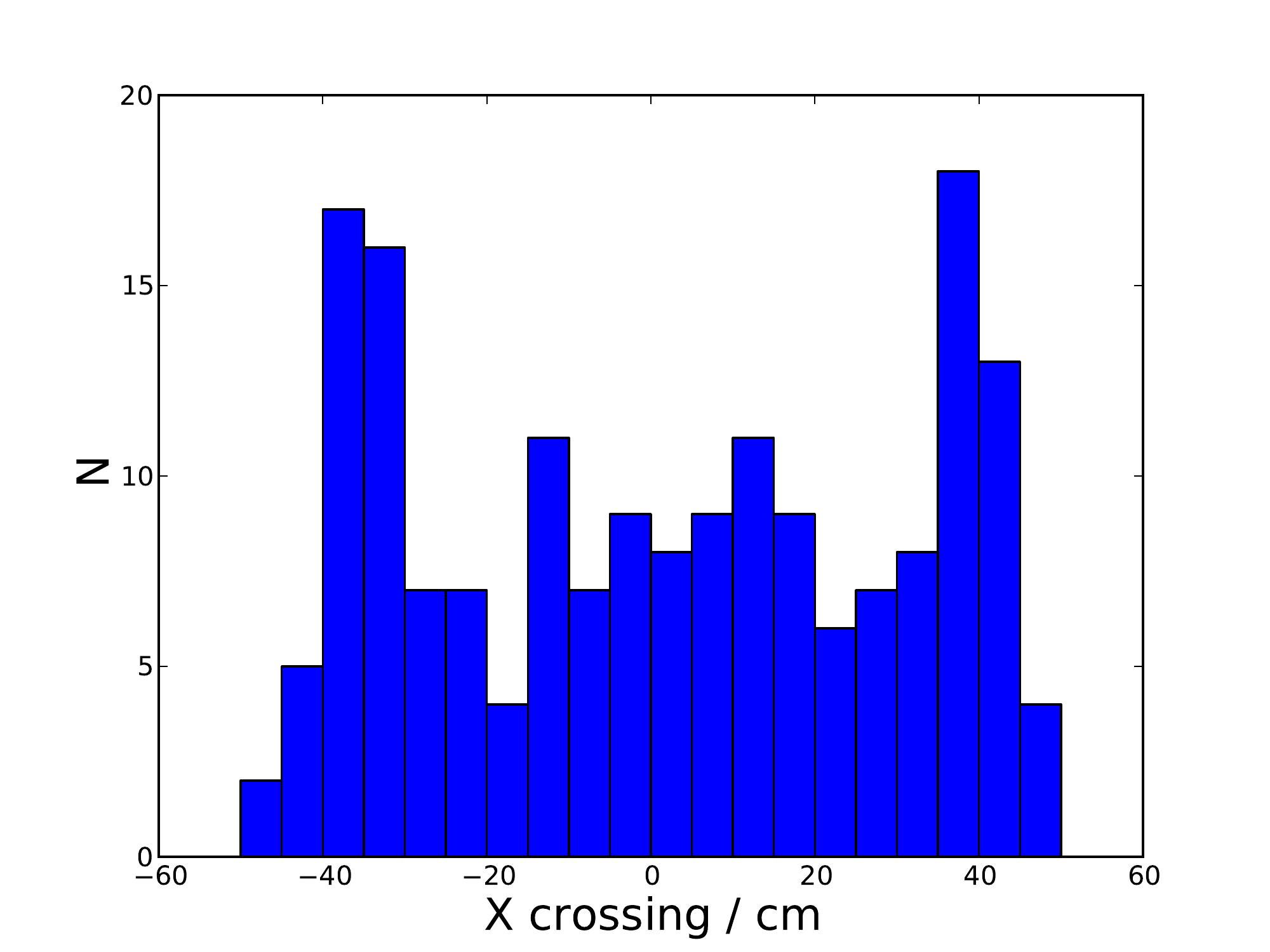}
      \label{fig:rhov_l_400}
   }
   \caption{Crossing points for the individual pedestrians, showing the effective width of the corridor.}
   \label{fig:effectivewidth_l}
\end{figure}

\begin{figure}[ht]
   \centering
   \subfigure[$l = 6$ cm]
   {
      \includegraphics[scale=0.4]{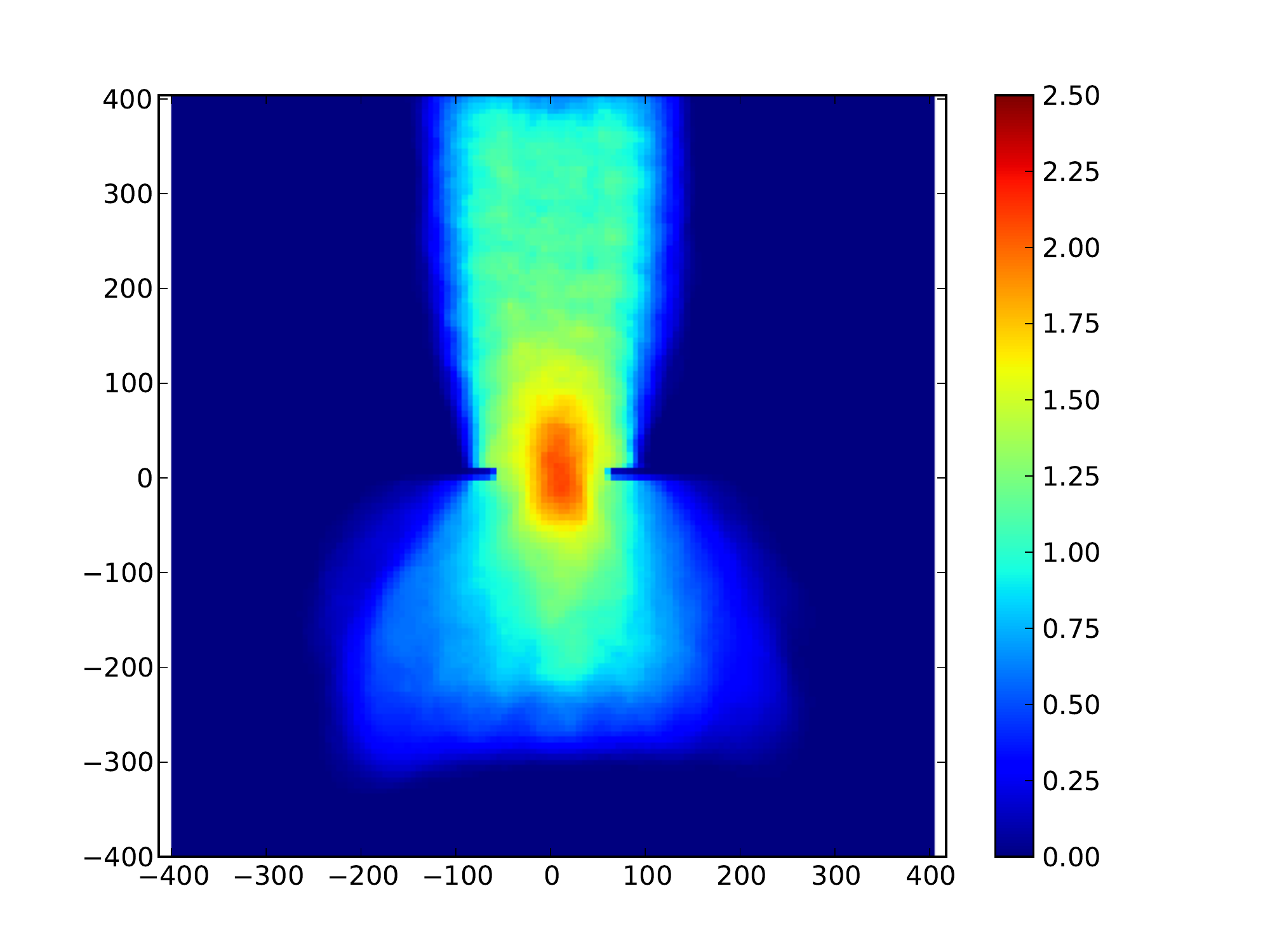}
      \label{fig:rhov_l_006}
   }
   \subfigure[$l = 200$ cm]
   {
      \includegraphics[scale=0.4]{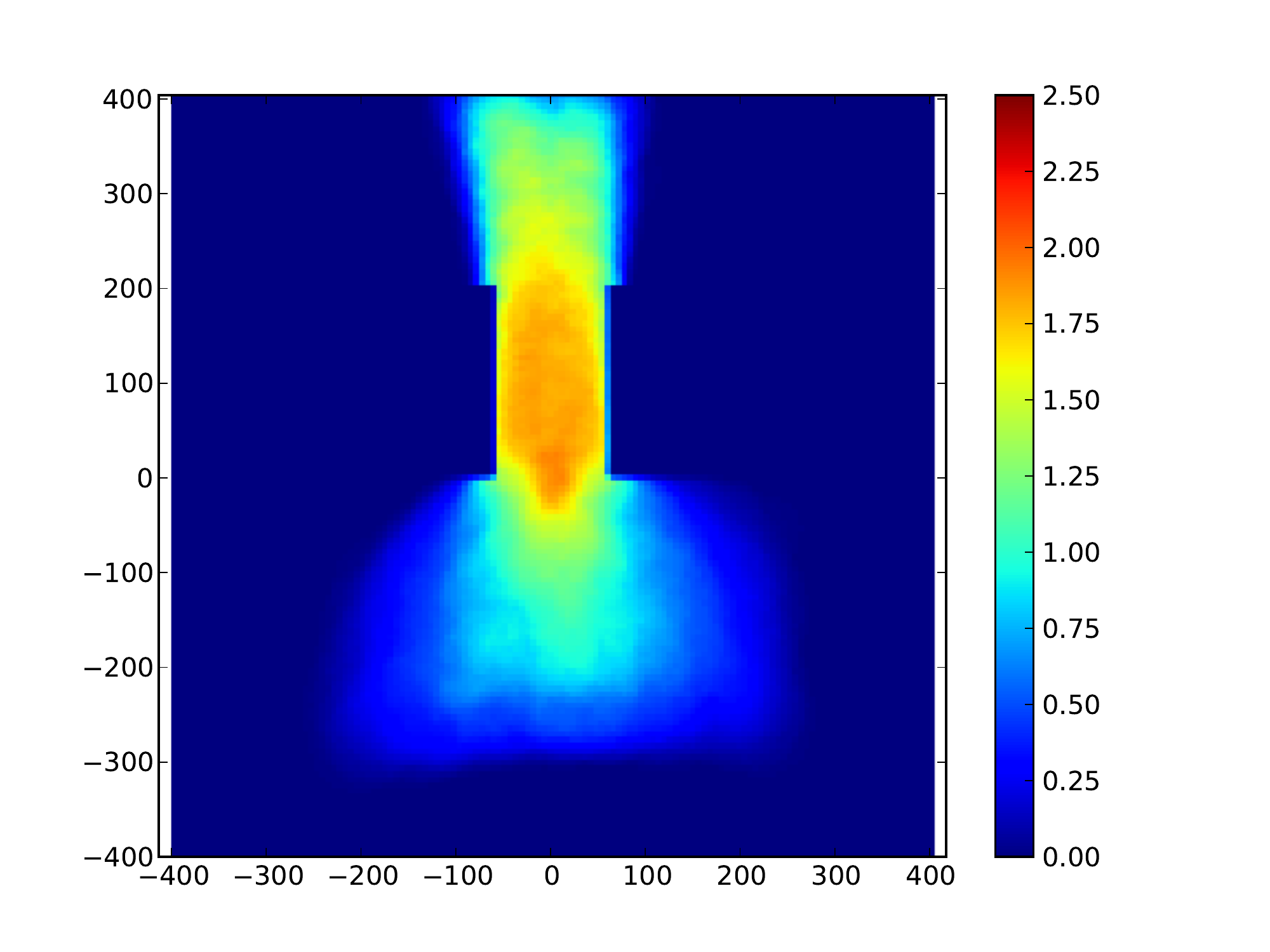}
      \label{fig:rhov_l_200}
   }
   \subfigure[$l = 400$ cm]
   {
      \includegraphics[scale=0.4]{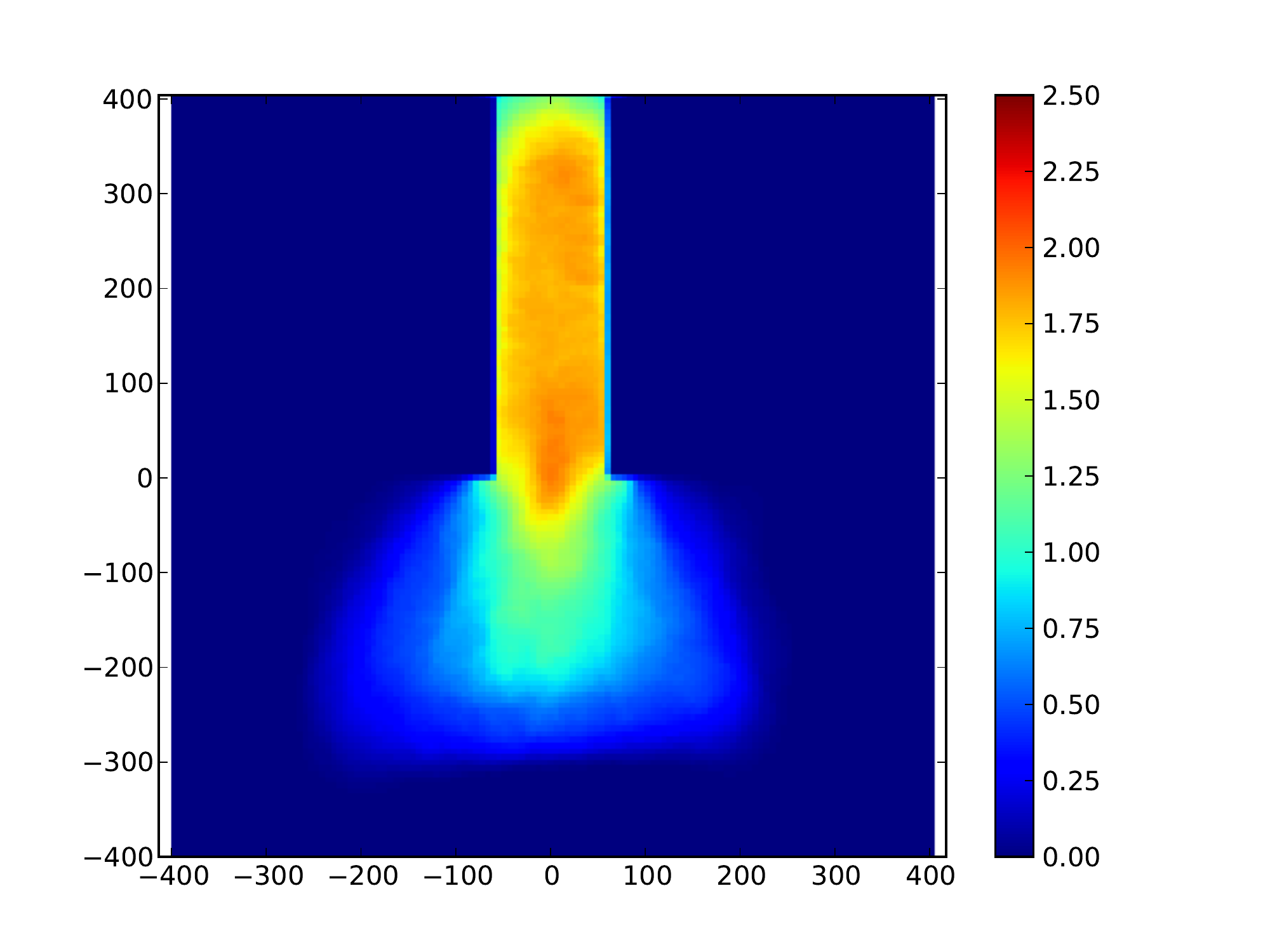}
      \label{fig:rhov_l_400}
   }
   \caption{Maps of the specific flow, $J_s$.}
   \label{fig:rhov_l}
\end{figure}

\begin{figure}[ht]
   \centering
   \subfigure[$l = 6$ cm]
   {
      \includegraphics[scale=0.4]{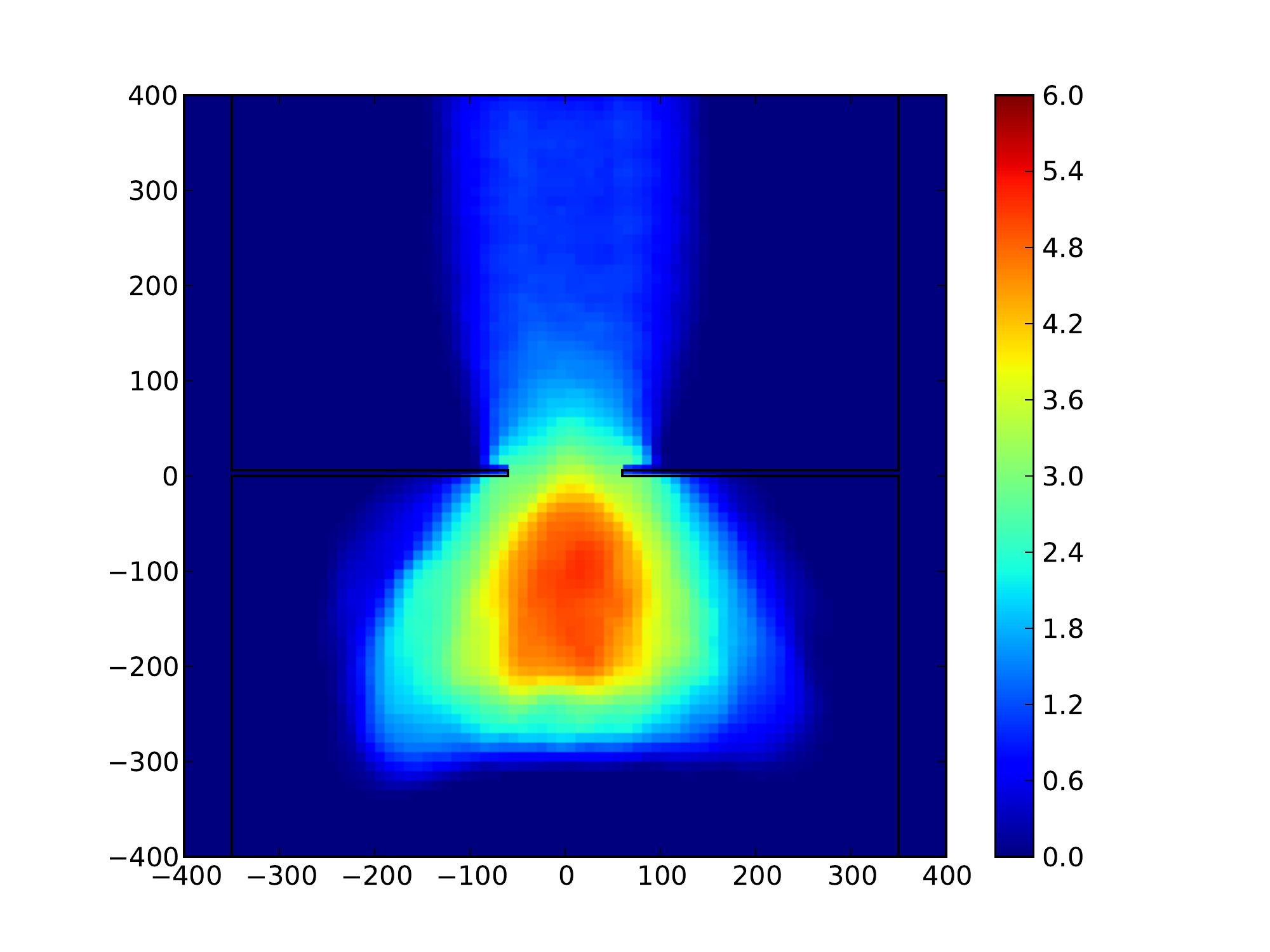}
   }
   \subfigure[$l = 200$ cm]
   {
      \includegraphics[scale=0.4]{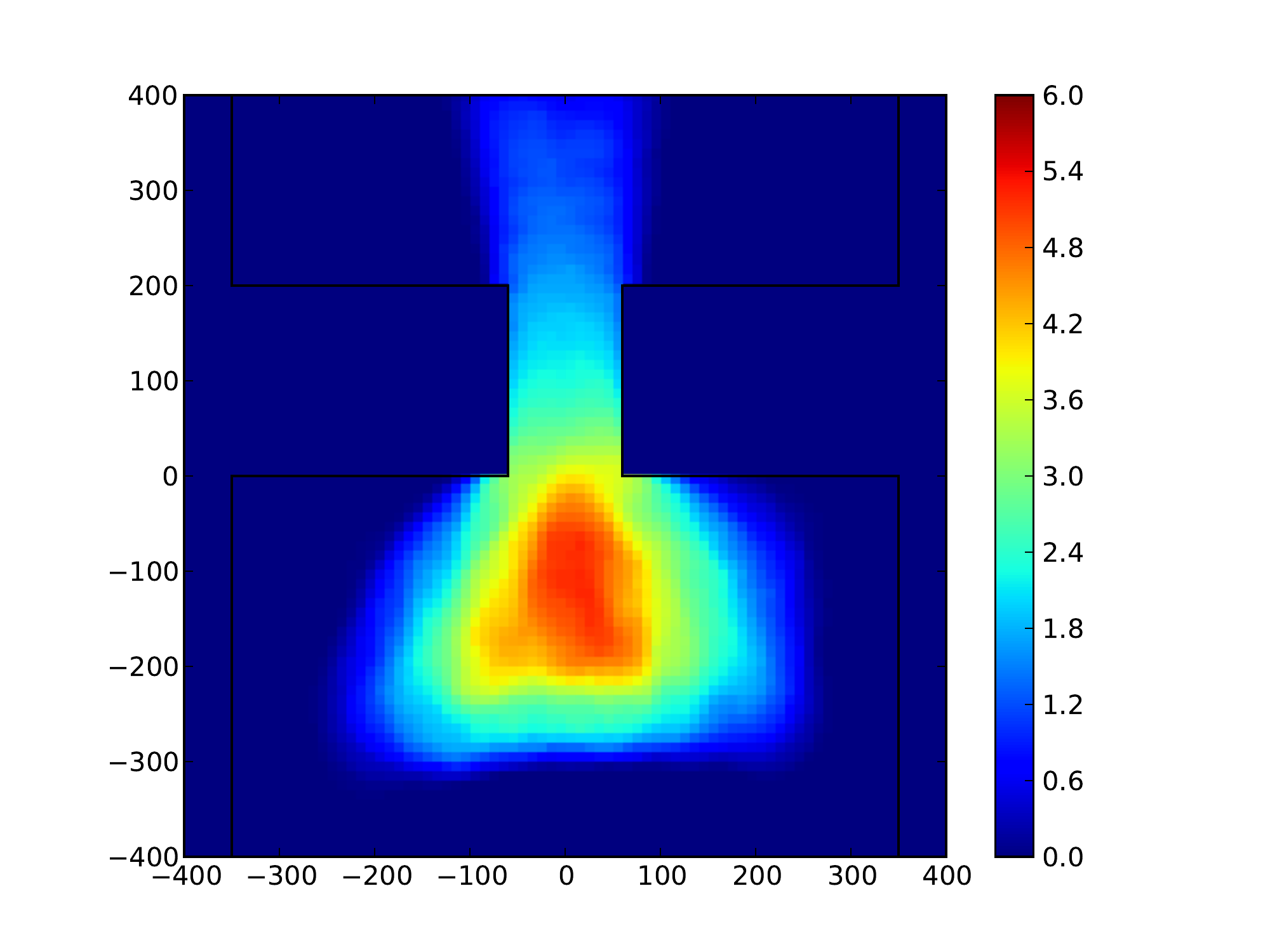}
   }
   \subfigure[$l = 400$ cm]
   {
      \includegraphics[scale=0.4]{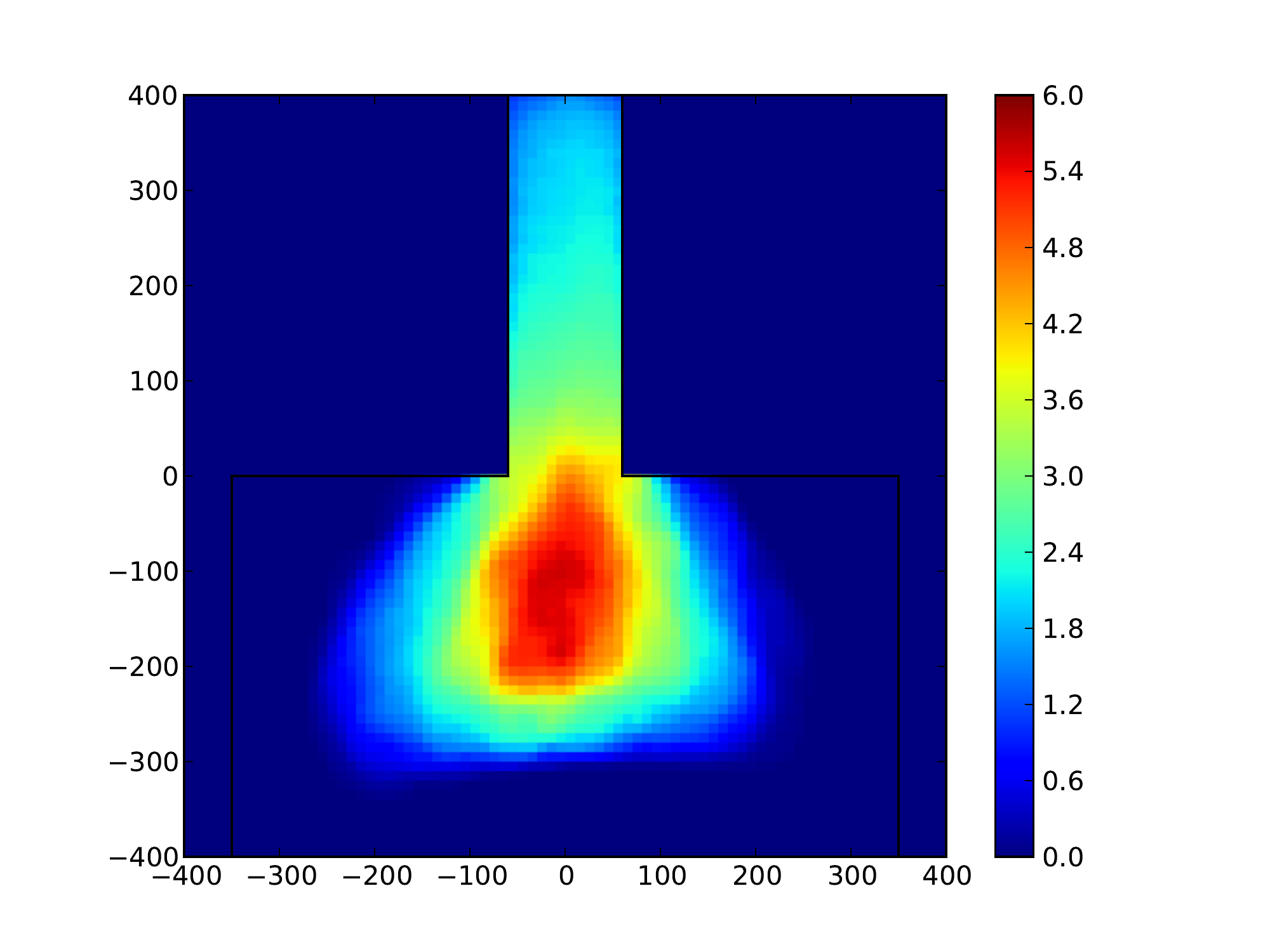}
   }
   \caption{Maps of the integrated density.}
   \label{fig:d_l}
\end{figure}

\begin{figure}[ht]
   \centering
   \subfigure[$l = 6$ cm]
   {
      \includegraphics[scale=0.4]{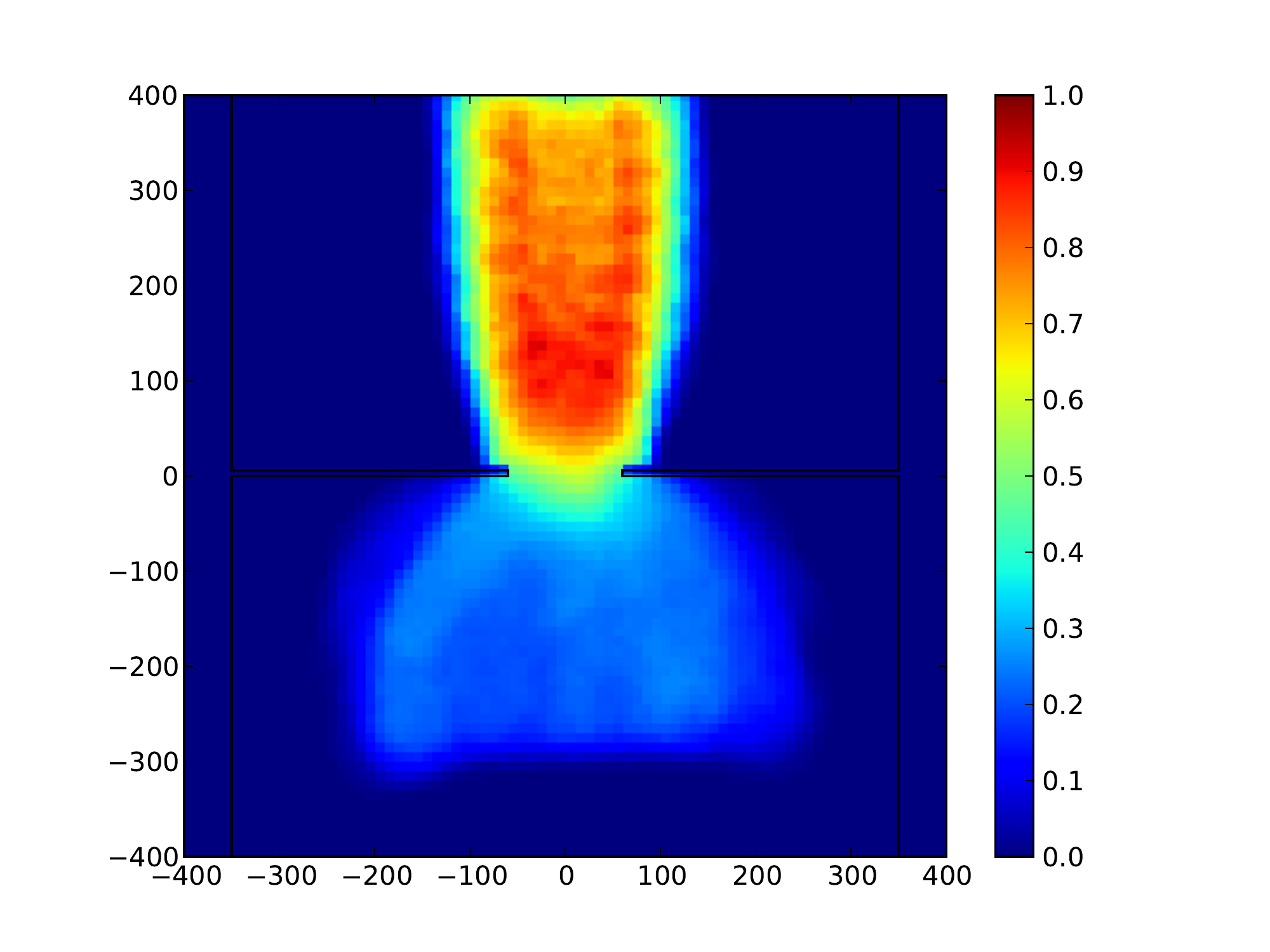}
   }
   \subfigure[$l = 200$ cm]
   {
      \includegraphics[scale=0.4]{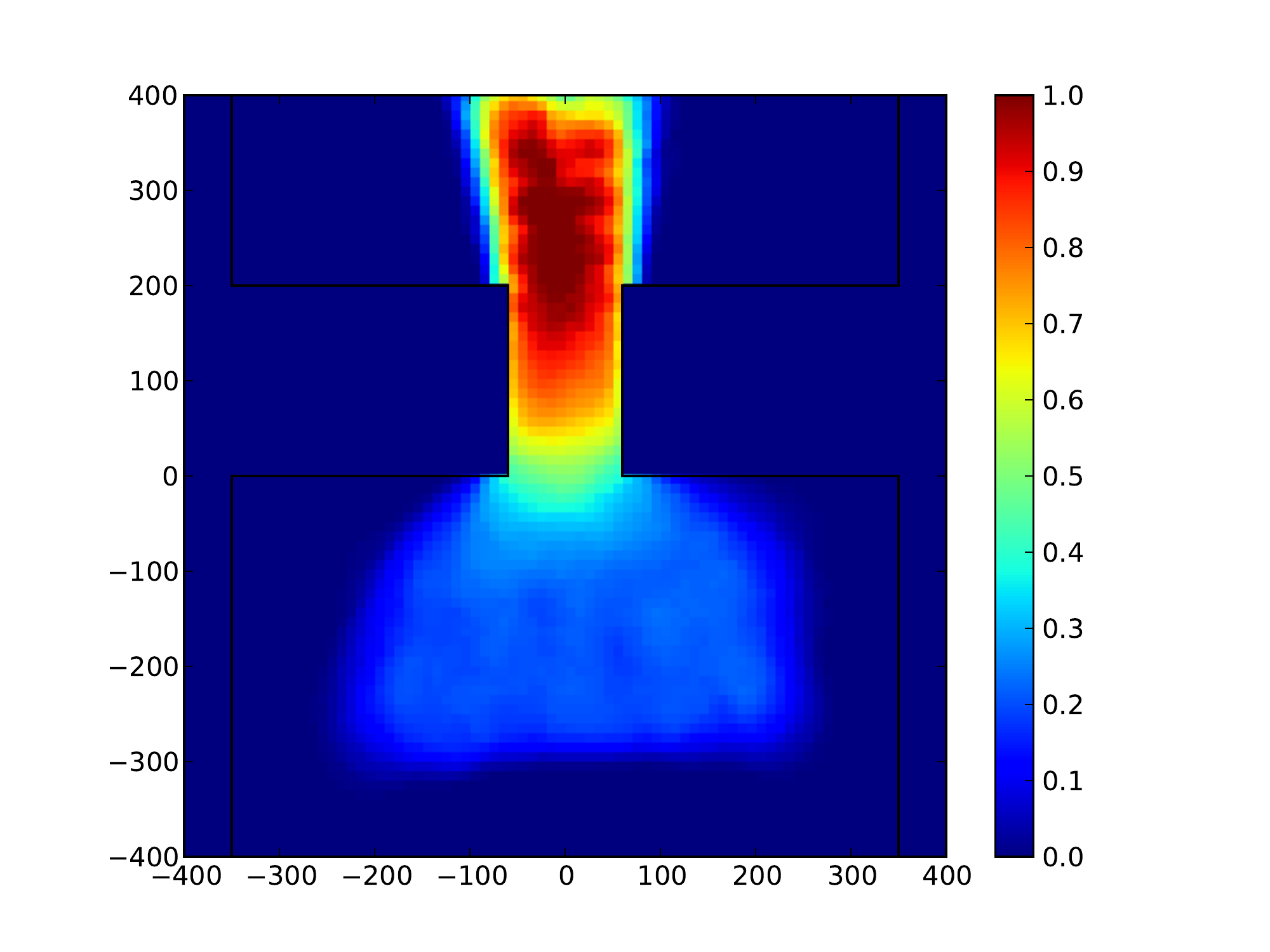}
   }
   \subfigure[$l = 400$ cm]
   {
      \includegraphics[scale=0.4]{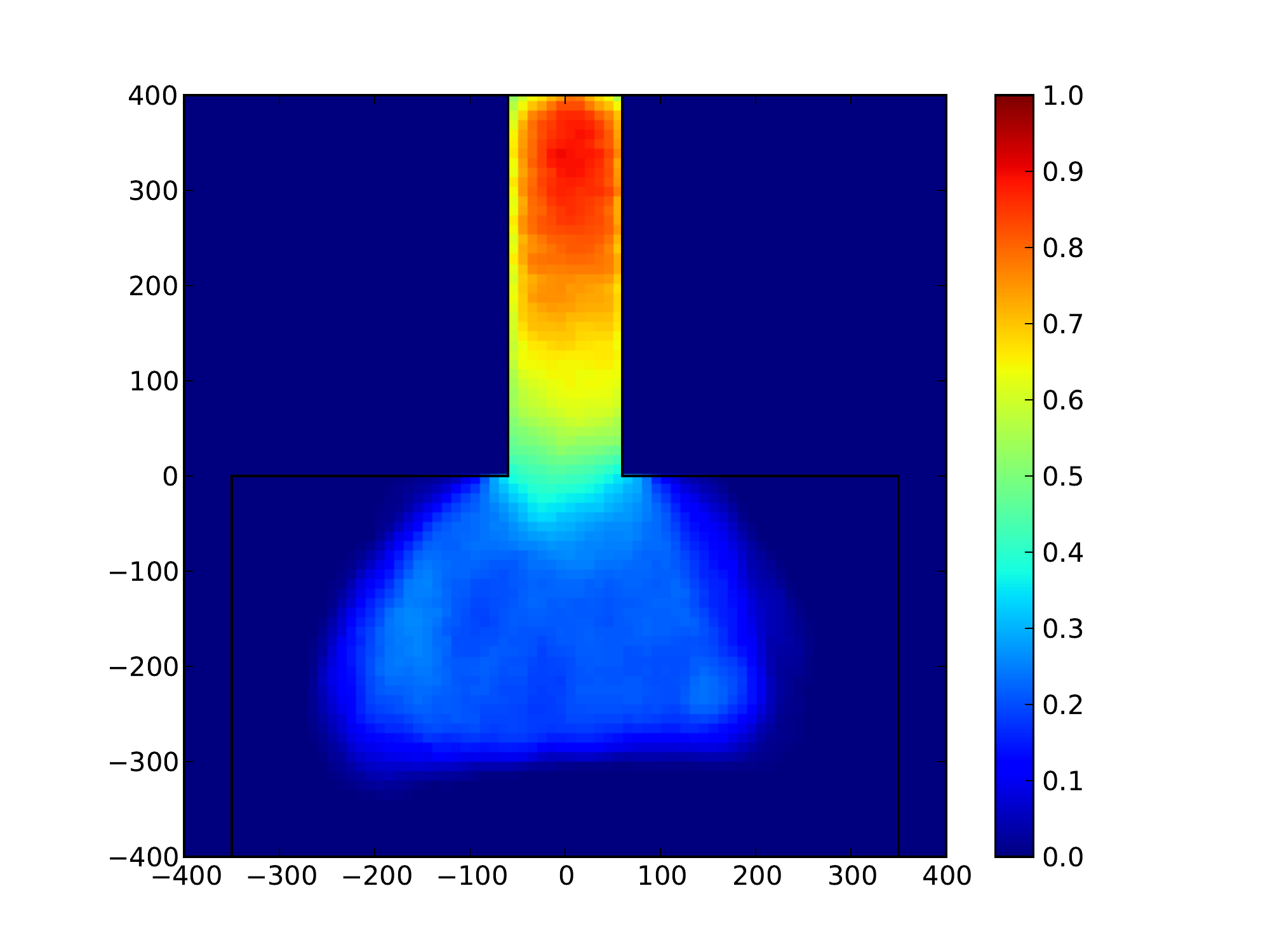}
   }
   \caption{Maps of the integrated velocity.}
   \label{fig:v_l}
\end{figure}

\begin{figure}[ht]
   \centering
   \includegraphics[scale=0.3]{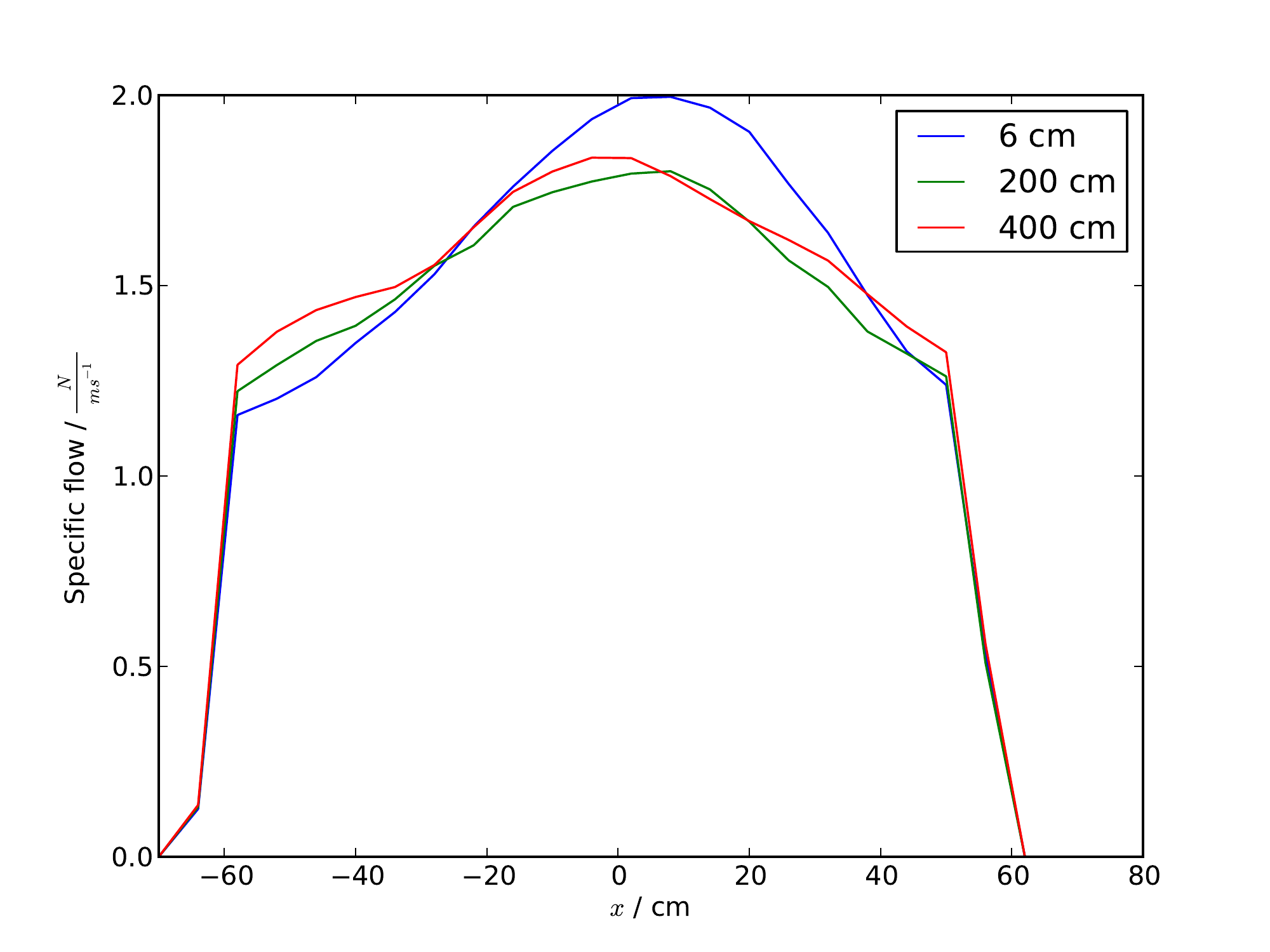}
   \caption{Specific flow measured over 10 cm patches.  Slice taken through bottleneck entrance.}
   \label{fig:rhov_slice}
\end{figure}

\begin{figure}[ht]
   \centering
   \includegraphics[scale=0.3]{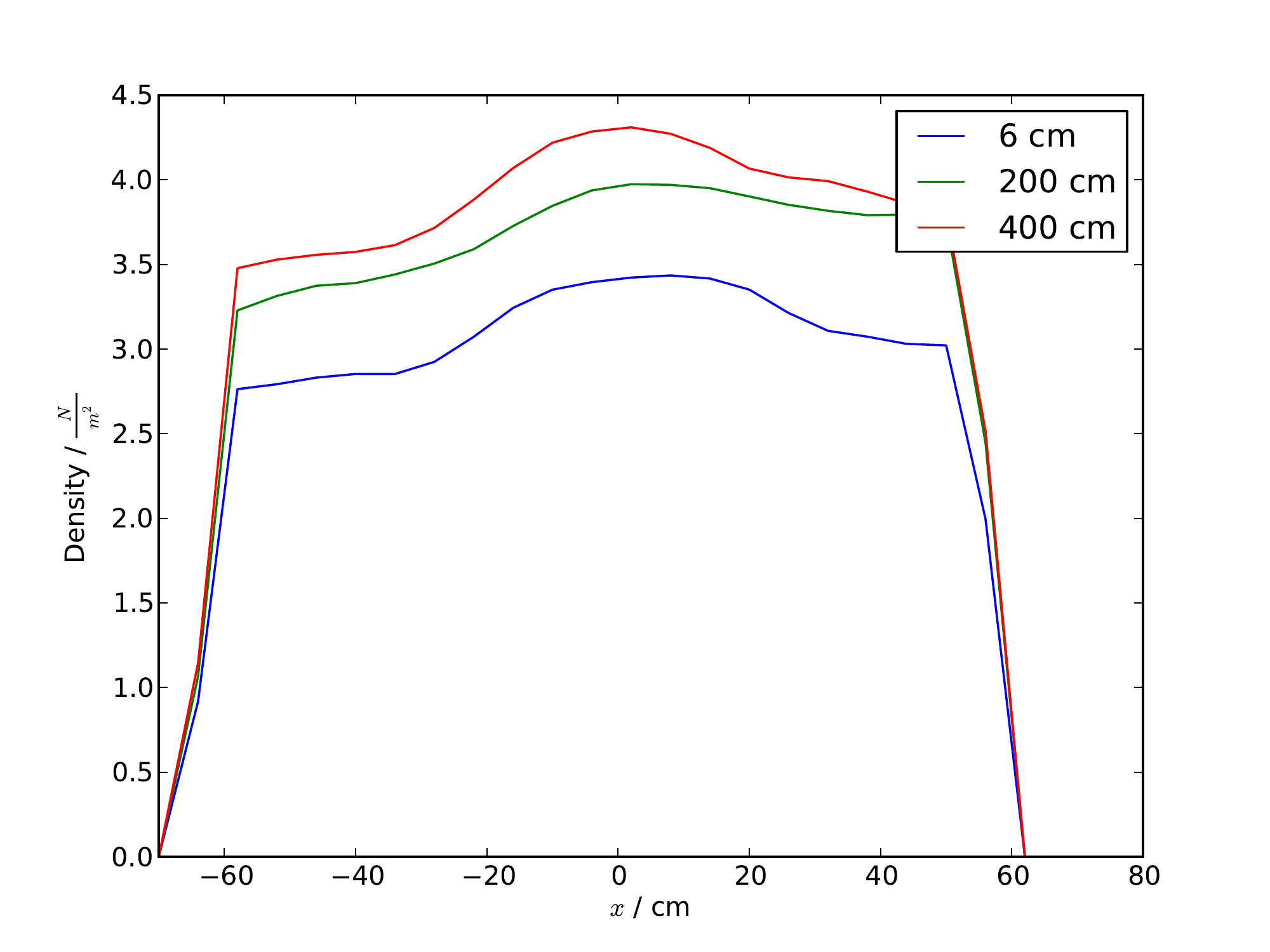}
   \caption{Integrated density measured over 10 cm patches.  Slice taken through bottleneck entrance.}
   \label{fig:dens_slice}
\end{figure}

\begin{figure}[ht]
   \centering
   \includegraphics[scale=0.3]{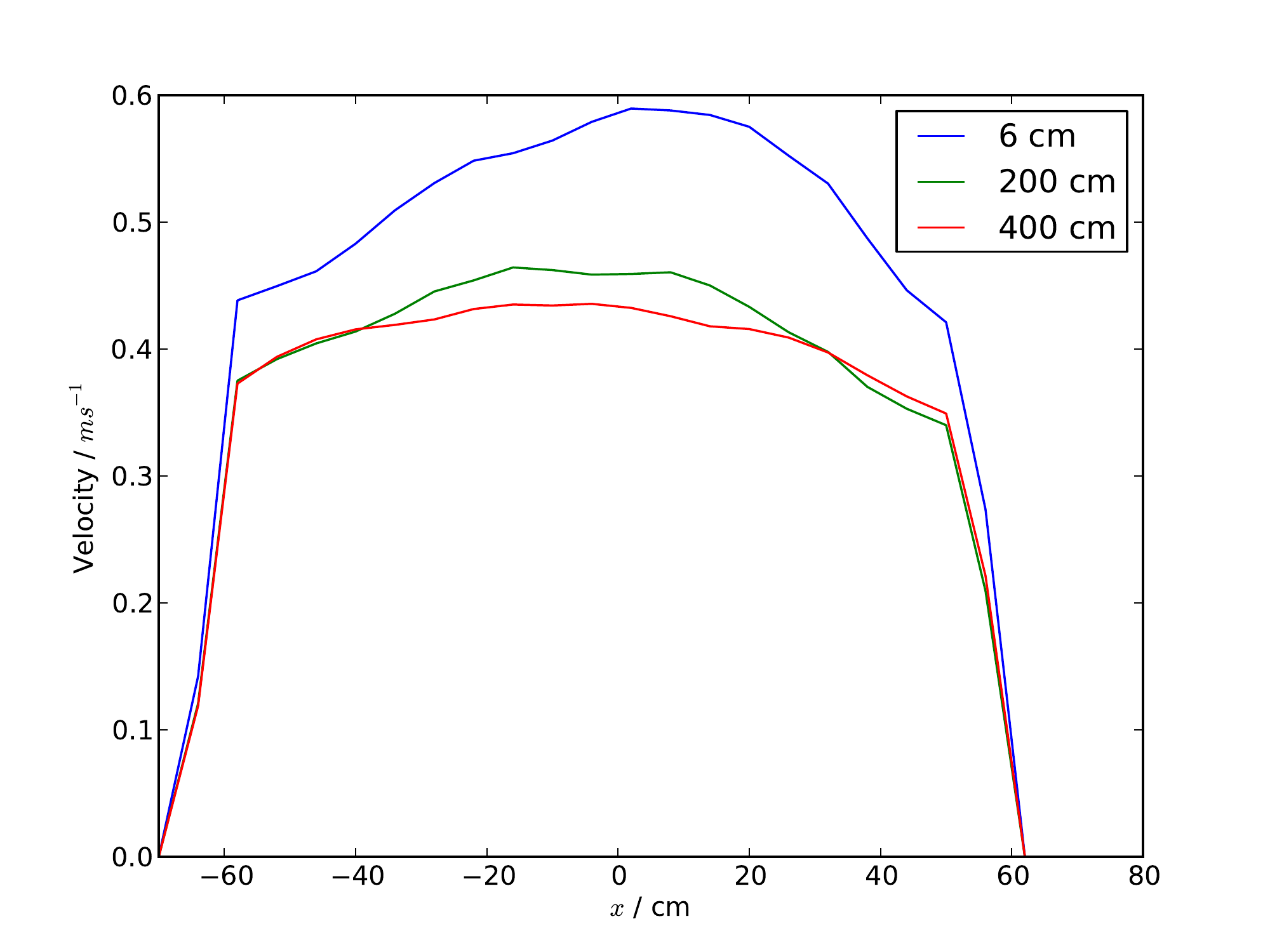}
   \caption{Integrated velocity flow measured over 10 cm patches.  Slice taken through bottleneck entrance.}
   \label{fig:vels_slice}
\end{figure}

We now must ask ourselves why this happens.  Is the increased flow due to greater pedestrian density at the bottleneck entrance or due to increased velocity?  
In figure~\ref{fig:effectivewidth_l} the horizontal position where the pedestrians enter the bottleneck is shown.  These histograms show that there is no noticeable increase in the effective width for the shortest bottleneck.  The maps (figures~\ref{fig:d_l} and~\ref{fig:v_l}) and accompanying slices (figures~\ref{fig:dens_slice} and~\ref{fig:vels_slice}) demonstrate that the density at the entrance to the short bottleneck is significantly lower and the velocity significantly higher.  Pedestrians passing through the short bottleneck can side step the barrier, in a single pace.  This takes them from infront of the bottleneck where their motion is restricted by the walls and other pedestrians, to outside the bottleneck where their motion is significantly less restricted and they are free to move in any direction they choose.  This act of side-stepping is responsible for the lowered density at the entrance to the short bottlenecks, and the freer motion responsible for the increased velocity.  

\clearpage
\section{Conclusions}
We have shown that the integrated density possess many advantages over the classical definition.  These are the greatly improved temporal resolution and the ability to obtain information about the system over microscopic scales.

The reduced fluctuations showed us the non-stationarity in the density for the narrowest bottlenecks.  We do not attribute this non-stationarity to habituation to the experimental setup, but instead to the emergence of a physical effect.  This is an important result that merits further investigation, if reproducible it implies that in many real applications the density is not constant over time.  The microscopic analysis reveals facets of the density and velocity distributions which cannot be obtained using standard methods.  The density peak moves into the bottleneck as the bottleneck widens.  The integrated density is in agreement with the standard definition, systematic differences between the methods can be understood in terms of the scale sensitivity of the methods.

The anomalous flow through the short bottlenecks was considered it has been shown that the flow becomes stationary, and that the flow for the short bottleneck is significantly wider.  The profile of the specific flow studied and the increased specific flow was attributed to the presence of a new stepping mode, only permitted for short bottlenecks.
\bibliographystyle{unsrt}
\bibliography{voronoi_paper}

\end{document}